\documentclass[11pt]{article}
\pdfoutput=1
\usepackage[utf8]{inputenc}
\usepackage{braket,slashed,bm}
\usepackage{jheppub} \usepackage{braket,slashed,bm}
\usepackage{array,multirow}
\usepackage{amsmath}
\usepackage[normalem]{ulem}
\usepackage[T1]{fontenc}
\usepackage{mathrsfs}
\usepackage{simplewick}
\usepackage{tikz}
\newcommand{\abs}[1]{\left|#1\right|}
\usepackage{subcaption}
\usetikzlibrary{arrows,decorations.pathmorphing,backgrounds,positioning,fit,petri,automata,shadows,calendar,mindmap,
decorations.markings,calc}
\usepackage{feynmp-auto}
\usepackage{units}
\def\beq{\begin{equation}}
\def\eeq{\end{equation}}
\def\bea{\begin{eqnarray}}
\def\eea{\end{eqnarray}}
\def\nn{\nonumber \\}

\renewcommand{\a}{\alpha}
\renewcommand{\b}{\beta}
\renewcommand{\d}{\delta}

\newcommand{\s}{\sigma}
\newcommand{\de}{\partial}

\newcommand{\Lagr}{\mathcal{L}}
\renewcommand{\a}{\alpha}
\renewcommand{\b}{\beta}
\newcommand{\des}{\slashed{\partial}}
\newcommand{\hc}{{\rm h.c.}}
\newcommand{\diag}{{\rm diag}}
\renewcommand{\Re}{{\rm Re}}
\renewcommand{\Im}{{\rm Im}}
\newcommand{\DistTo}{\xrightarrow{
   \,\smash{\raisebox{-0.65ex}{\ensuremath{\scriptstyle\sim}}}\,}}

\title{Examining the neutrino option}
\author{
Ilaria Brivio and Michael Trott\\
Niels Bohr International Academy and Discovery Centre, Niels Bohr Institute,
University of Copenhagen, Blegdamsvej 17, DK-2100 Copenhagen, Denmark}
\abstract{The neutrino option is a scenario where the electroweak scale, and thereby the Higgs mass, is generated simultaneously with neutrino masses in the seesaw model. This occurs via the leading one loop and tree level diagrams matching the seesaw model onto the Standard Model Effective Field Theory. We advance the study of this scenario by determining one loop corrections to the leading order matching results systematically, performing a detailed numerical analysis of the consistency of this approach with Neutrino data and the Standard Model particle masses, and by examining the embedding of this scenario into a more ultraviolet complete model. We find that the neutrino option remains a viable and intriguing scenario to explain the origin of observed particle masses.}
\begin{document}
\maketitle

\section{Introduction} \label{sec:intro}
The origin of the Higgs potential and the observed neutrino masses are two outstanding questions unanswered by the Standard Model (SM). The smallness of neutrino masses compared to the masses of the quarks and Electroweak gauge bosons can be naturally accommodated
in the Seesaw model \cite{Minkowski:1977sc,GellMann:1980vs,Mohapatra:1979ia,Yanagida:1980xy,Schechter:1980gr}.
In this approach, the neutrino-quark mass hierarchy follows from a separation of the scale (denoted $M$) associated with a lepton number violating sector
leading to neutrino masses, compared to the Electroweak scale ($\bar{v}_T \equiv \sqrt{2 \langle H^\dagger H \rangle}$) associated with the
remaining SM field content.

If the Seesaw model generates neutrino masses, the sensitivity
of the Higgs mass term to $M$ can lead to a  curious tuning of Lagrangian parameters.
Such parameter tuning can be avoided -- if
the Majorana mass scale and couplings are of a particular form \cite{Vissani:1997ys,Brivio:2017dfq}.
For the same region of parameter space, the Seesaw model can be a phenomenologically viable boundary condition for the Higgs potential,
while supplying an origin for the Electroweak scale as a descendent from the scale $M$ \cite{Brivio:2017dfq}.
What leads to the Higgs phase of the SM in this case, is Fermi statistics in the leading one loop matching contribution from the seesaw model
to the SM potential terms. This matching necessarily occurs while Neutrino masses are generated by the leading tree level matching, as the Majorana states are integrated out.
This is the ``neutrino option'' for generating neutrino masses
and the Higgs potential.

The purpose of this paper is to examine
this scenario in more numerical and theoretical detail than Ref.~\cite{Brivio:2017dfq}
and to illustrate how this scenario can be embedded in more concrete Ultraviolet (UV) models.\footnote{As this paper was being drafted,
the neutrino option was embedded in a classically conformal UV physics scenario by  Brdar et al \cite{Brdar:2018vjq}. We
discuss this interesting proposal in some detail and extend it to include gravitational interactions.}
We consider the minimal case that can lead to a successful lower energy neutrino phenomenology with two heavy Majorana states leading to two massive light neutrinos.

The outline of this paper is as follows. In Section \ref{sec:framework} we define our notation and conventions
for the Seesaw model and the leading tree level matching onto the Standard Model Effective Field Theory (SMEFT). In Section \ref{sec:NLOanalysis}
we develop the theoretical framework to study the neutrino option at next to leading order (NLO) accuracy, determining the relevant
one loop matching into the seesaw model, and discuss the one loop running of the neutrino parameters. In Section \ref{sec:numerics} we report numerical results fixing the low scale neutrino parameters  and Higgs mass as inputs,
and determine the $\lambda$ parameter required at the matching scale for the scenario to be self consistent.
In Section \ref{sec:UV} we examine and extend the embedding of the neutrino option into a classically (i.e. massless particle spectrum at leading order) conformal UV completion, as recently proposed by Brdar et al \cite{Brdar:2018vjq}. We discuss the cut off scale in this particular UV embedding and the potential to avoid fine tuning.
We briefly comment on the possibility that this UV framework can simultaneously provide a Dark Matter candidate, and in Section \ref{sec:conclusions} we conclude.

\section{Theoretical framework}\label{sec:framework}

The physics of the SM is such that global lepton number conservation can provide an accidental symmetry protection mechanism for the Higgs potential terms.
The operator dimension in the SM, or the SMEFT operator expansion
leads to the association of an operators dimension being even (odd) if
$(\Delta B - \Delta L)/2$ is even (odd)\cite{deGouvea:2014lva,Kobach:2016ami} due to the nature of the SM field content. Here $\Delta B$
and $\Delta L$ are the baryon and lepton number violation of the operator considered.
$(H^\dagger H)$ is of even dimension with $\Delta B = \Delta L =0$.
Lepton number carrying fields, associated with the mass scale $M$, either couple in pairs to $H^\dagger H$, or
with a dimensionful coupling expected to be proximate to $M$, if parameter tuning is avoided.
Tree level exchanges leading to $(H^\dagger H)^2$ then have a cancellation
of the introduced $M$ dependence (up to ratios of couplings). The coupling of lepton number violating fields also takes place
through the portal interaction $H^\dagger \ell$ (+ h.c.). A heavy scale associated with lepton number violating fields
 leads to an inverse dependence on this scale at tree level, starting with the ($\Delta L = 2$)
dimension five Weinberg operator, due to these interactions\footnote{Note that in models with multiple BSM scales, such as the well-known inverse seesaw model~\cite{Mohapatra:1986aw,Mohapatra:1986bd,Bernabeu:1987gr}, it is possible to make $M$ distinct from the quantity that controls the lepton number violation. In this case the latter can be naturally very small and suppress the Wilson coefficients multiplicatively, while allowing lower values of $M$ ($\sim$ TeV).}.  The minimal scenario of this form is the Seesaw model of neutrino mass generation.
An expectation is that small neutrino masses result, made only smaller by
any small couplings $(\omega)$ to heavy particles. Simultaneously, {\it additive contributions} to the Higgs mass
parameter appear at one loop $ \propto \omega^2 M^2/16\pi^2$. These loop corrections are not forbidden, as lepton number
is an accidental symmetry.

The basic pattern
of mass scales associated with the Electroweak scale, Higgs mass and neutrino masses can then be
\bea
m_\nu \sim \frac{\omega^2 \, \bar{v}_T^2}{M}, \quad \quad m_h \sim \frac{\omega M}{4 \pi}, \quad \quad \bar{v}_T \sim \frac{\omega M}{4 \, \sqrt{2} \, \pi \, \sqrt{\lambda}},
\eea
and the Seesaw model parameter space $M \sim 10^7 \, {\rm GeV}$ and $|\omega|\sim 10^{-4}$, which leads to $m_\nu  \sim 10^{-11} \, {\rm GeV}$ and $m_h \sim 10^2  \, {\rm GeV}$, is particularly
interesting.\footnote{For related results see Ref.~\cite{Davoudiasl:2014pya,Casas:1998cf,Casas:1999cd,Bambhaniya:2016rbb}.  }

This is the pattern of masses expected when the neutrino option is used to generate the Higgs potential.
The idea is to use a UV boundary condition to generate the Higgs potential and an {\it effective Electroweak scale}.
The smallness of the Higgs mass parameter is linked to the small neutrino masses and a
set of approximate symmetries: global lepton number and scale invariance.
For the latter, an expansion around the approximately scale invariant limit of the SM\footnote{In this work we use scale invariance and conformal invariance interchangeably as we are considering tree level effects of this symmetry,
see Refs.~\cite{Callan:1970ze,Coleman:1970je} for foundational discussions. Ref.~\cite{Bardeen:1995kv} is the first reference, to our knowledge, discussing the use of scale invariance to address the Hierarchy problem. Scale invariance is of course anomalous \cite{Capper:1974ic}, but still useful to consider in this manner.}, incorporating the soft breaking of approximate scale invariance feeding into
the Higgs potential due to the scale $M$ is done. (In the limit $M \rightarrow 0$, $\langle H^\dagger H \rangle \DistTo \Lambda_{QCD}^2$.)
We stress however, that the excitation of the Higgs field is not the dilaton of spontaneously broken scale invariance in this approach.

The motivation to consider this possible origin of the Electroweak scale is largely supplied by current experimental results.
Neutrino's are known to be massive states, requiring an extension of the SM.  It is natural to consider the effects of extending the SM
to generate Neutrino masses on the Higgs. Generating the Higgs potential around the scales probed by LHC, with partner states associated with the multiplets of an
approximate stabilizing symmetry (such as SUSY), or through
lowering the effective Planck scale, is now subject to
increasingly severe experimental bounds.\footnote{For a good discussion on the theoretical challenges of generating the Higgs potential around the
$\sim {\rm TeV}$ scale in composite models, see Ref.~\cite{Bellazzini:2014yua}.} Conversely, the neutrino option uses the running of the Higgs potential parameters
in conjunction with threshold matchings required for Neutrinos to have mass when generating the Higgs potential. This occurs at scales far above the observed Higgs mass
and generates an effective Electroweak scale with scant experimental evidence of any stabilizing symmetry.\footnote{In such a scenario, technical fine tuning can be avoided, while new states are absent at the LHC. It is also possible that an experimental
signature of $\sim {\rm GeV} -{\rm TeV}$ new scalar states could exist in exceptional regions of parameter space. We discuss this possibility in Section \ref{sec:darkmatter}.}
The advantage of this approach is that a simple spectra of new physics states, motivated out of the experimental fact that neutrino's are massive
can lead to the observed "mexican-hat" potential at lower scales, once the Higgs potential is run down using the Renormalization
Group Equations (RGE) of the SM.  A bare value of $\lambda_0 \sim 0.01$ is required, in addition to the threshold matching contribution to $\lambda$ from the seesaw model. This is the main result of this paper's numerical study. As the Higgs is not a pseudo-Goldstone boson of any symmetry -- a bare $\lambda$ parameter is also not forbidden.

\subsection{The SMEFT}
We study the neutrino option in the SMEFT \cite{Buchmuller:1985jz,Grzadkowski:2010es},
where the SM is extended with higher dimensional operators to capture the low energy limit of the seesaw model
\begin{align}
	\Lagr_{\textrm{SMEFT}} &= \Lagr_{\textrm{SM}} + \Lagr^{(5)}+\Lagr^{(6)} +
	\Lagr^{(7)} + \dots, &  \quad \quad  \Lagr^{(d)} &= \sum_i \frac{C_i^{(d)}}{M^{d-4}}\mathcal{Q}_i^{(d)},
	\textrm{ for } d>4.
\end{align}
Here  $Q_i^{(\rm{d})}$ are suppressed by $\rm{d}-4$ powers of the Majorana scale $M$, that acts as the cut-off, and
the $C_i^{(\rm{d})}$ are the Wilson coefficients. Our SM notation is consistent with the SMEFT review \cite{Brivio:2017vri} except for the modified notation for the Higgs potential terms
\bea
 V(H^\dag H) &=& -\frac{m^2}{2}(H^\dag H) + \lambda (H^\dag H)^2 + \cdots,  \nn
 &=& -\frac{m_0^2 + \Delta m^2}{2}(H^\dag H) + (\lambda_0 + \Delta \lambda) (H^\dag H)^2 + \cdots.
\eea
$\widetilde H_j = \epsilon_{jk} (H^{k})^\star$ and the star superscript is generally reserved for
complex conjugation on bosonic quantities.
Here the bare parameters are $m_0, \lambda_0$.
In a classical conformal limit for the mass spectrum of the SM (with non conformal renormalized couplings) $m_0^2 \simeq 0$ while $\lambda_0$ is unconstrained.
For the Weinberg operator \cite{Weinberg:1979sa,Wilczek:1979hc} we use the notation
\bea\label{weinberg}
\mathcal{Q}_{\alpha \beta}^{(5)} = \left(\overline{\ell^{c, \alpha}_L} \, \tilde{H}^\star\right) \left(\tilde{H}^\dagger \, \ell^{\beta}_L\right).
\eea
Our spinor conventions are that the $c$ superscript corresponds to a charge conjugated Dirac four-component spinor $\psi^c  = C \overline{\psi}^T$
with $C= - i \gamma_2 \, \gamma_0$ in the chiral basis for the $\gamma_i$ we use.  Chiral projection and $c$ do not commute so we fix notation that
$\ell^c_L$ denotes a doublet lepton field chirally projected and subsequently charge conjugated.

\subsection{Seesaw model}
We use the notation and conventions of Refs.~\cite{Broncano:2002rw,Elgaard-Clausen:2017xkq} for the Seesaw model.
In the Seesaw model, the SM Lagrangian field content is extended with right handed singlet fields $N_{R,p}$ with
vanishing $\rm SU(3) \times SU(2)_L \times U(1)_Y$ charges. As these are singlet fermion fields they have Majorana mass terms
\cite{Majorana:1937vz} of the form
\bea
\overline{N_{R,p}^c} \, M_{pr} \, N_{R,r} + \overline{N_{R,p}} \, M^\star_{pr} \, N_{R,r}^c,
\eea
where the charge conjugate of $N_R$ is $N_{R}^c$. We define a field satisfying the Majorana condition as $N_p = N_p^c$
in its mass eigenstate basis as \cite{Bilenky:1980cx,Broncano:2002rw}
\bea
N_p = e^{i \theta_p/2} \, N_{R,p} + e^{- i \theta_p/2} \, (N_{R,p})^{c}.
\eea
With this choice, all Majorana phases $\theta_p$ shifted into the effective couplings and
the relevant terms in the UV Lagrangian are
\begin{equation}\label{L2formulation}
 \mathcal{L}_N = \frac{1}{2}\bar N_p (i\des-M_p)N_p -
 \frac{1}{2}\Bigg[ \overline{\ell_L^\beta}\tilde H \omega_\beta^{p,\dag}N_p + \overline{\ell_L^{c\beta}}\tilde H^* \omega_\beta^{p,T}N_p
 +\overline{N_p} \omega_\beta^{p,*}\tilde H ^T \ell_L^{c\beta} +\overline{N_p} \omega_\beta^p\tilde H^\dag \ell_L^{\beta}
 \Bigg]\,.
\end{equation}
Here $p=\{1,2\}$ runs over the heavy $N_p$ Majorana states ($M_p \sim M$), while $\beta=\{1,2,3\}$ runs over the SM lepton flavors.
This formulation of the Seesaw model is mathematically equivalent to the formulation where
\begin{equation}
 \mathcal{L}'_N = \frac{1}{2}\bar N_p (i\des-M_p)N_p -
\Bigg[ \overline{\ell_L^\beta}\tilde H \omega_\beta^{p,\dag}N_p  +\overline{N_p} \omega_\beta^p\tilde H^\dag \ell_L^{\beta}
 \Bigg].
\end{equation}
 In this case, the Lagrangian is reduced using the charge conjugation identities
and the Majorana condition for the field $N_p$.
Comparing calculations in these two formulations beyond tree level uncovers an interesting subtlety in using the Wick expansion, which is discussed in the Appendix.
\subsubsection{\texorpdfstring{$\mathcal{L}^{(5)}$}{L5} matching}
$\omega_{\beta}^p$ is a $\mathbb{C}_{2 \times 3}$ matrix, related to the physical light neutrino masses and mixings via matching onto the Weinberg operator
\begin{align}
 \mathcal{L}^{(5)} &= \frac{c^{(5)}_{\a\b}}{2} \mathcal{Q}_{\alpha \beta}^{(5)} +\hc, \quad    &
 c^{(5)}_{\a\b} &= \frac{(\omega^T)_{\a}^p \, \omega_\b^p}{M_p}.
\end{align}
Expanding the Higgs field around its classical background field gives
\begin{align*}
 \mathcal{L}^{(5)} &\supset -\frac{m_{\nu,k}}{2}\, \overline{\nu^{\prime c,k}_L} \nu_L^{\prime k}+\hc, \quad & {\rm where} \, \,& \quad &
 m_{\nu ,k} &= -\frac{v^2}{2} (U^T)_{k \a}\, c_{5,\a\b} \,U_{\b k},
\end{align*}
and $\nu_L^{\prime k}$ are the mass eigenstates of the light neutrinos $\nu_L^\a = U_{\a k}(\nu,L) \nu_L^{\prime k}$.
The matrix $U(\nu,L)$ rotates the neutrinos from their weak eigenstates to their mass eigenstates.
Similarly, the matrix $U(e,L)$ rotates between the charged lepton weak and mass eigenstates.
These rotations are not the same in general, leading to physical effects due to an overlap matrix being present in the lepton $\rm SU(2)_L$ doublet field $\ell$.
This is the Pontecorvo-Maki-Nakagawa-Sakata (PMNS) matrix \cite{Pontecorvo:1957cp,Maki:1962mu} defined as
\bea
\mathcal{U}_{PMNS} = \mathcal{U}^\dagger(e,L) \, \mathcal{U}(\nu, L).
\eea
The three eigenvectors in $U(e,L)$ form a basis for the field $\mathbb{C}^3$,
as they diagonalize $\mathcal{M}_e^\dagger \, \mathcal{M}_e$,  a Hermitian positive matrix defined over $\mathbb{C}^3$.
As the physical effects of the $\mathcal{U}_{PMNS}$ matrix come about due to the relative orientation of the eigenvectors defining
$U(e,L)$ and $U(\nu,L)$, we can choose the eigenvectors of $U(e,L)$ such that
$U(e,L) = {\rm diag}(1,1,1)$ as a basis for this space, so long as no physical conclusions depend on this choice. We use the parameterization
\begin{equation}\label{PMNS_def}
\begin{aligned}
U_{\rm PMNS} &\equiv V\cdot
                    \begin{pmatrix}
                     e^{-i \phi/2}& & \\ & e^{-i \phi'/2}& \\ & & 1
                    \end{pmatrix}, & \quad
V&=  \begin{pmatrix}
    c_2 c_3& s_3 c_2 & s_2 e^{-i\delta}\\
    -c_1 s_3-s_1 s_2 c_3 e^{i\delta} & c_1 c_3 -s_1 s_2 s_3 e^{i\delta}& s_1 c_2\\
    s_1 s_3 -c_1 s_2 c_3 e^{i\delta}& -s_1 c_3 - c_1 s_2 s_3 e^{i\delta}& c_1 c_2
    \end{pmatrix} \,,
\end{aligned}
\end{equation}
where $s_i=\sin\theta_i,\,c_i=\cos\theta_i$.

The matrix $m_{\nu ,k}$ has two non-zero eigenvalues in the case we consider. The lightest neutrino is massless, which is consistent with experimental results, and the mass eigenstates are labeled in descending order of mixing with the $\nu_e$ flavor eigenstate. With a Normal Hierarchy (NH) in masses one has then $0 = m_{\nu1}<m_{\nu2}<m_{\nu3}$, while for an
Inverted Hierarchy (IH) one has $0 = m_{\nu3}<m_{\nu1}<m_{\nu2}$. The two remaining masses are related to the squared mass differences
defined in Refs.~\cite{Gonzalez-Garcia:2014bfa,Esteban:2016qun} as\footnote{This notation can be related to that of the PDG with $\Delta m^2 = \Delta m_{3\ell}^2 \mp \Delta m_{21}^2/2$ for NH/IH.}
\begin{equation}
 \begin{cases}
 m_{\nu1} = 0\\
 m_{\nu2} =\sqrt{\Delta m_{21}^2} &({\rm NH})\\
 m_{\nu3} =\sqrt{\Delta m_{3\ell}^2} \\
 \end{cases}
 \qquad\qquad
 \begin{cases}
 m_{\nu3} = 0\\
 m_{\nu1} = \sqrt{-\Delta m_{21}^2 -\Delta m_{3\ell}^2}& ({\rm IH})\\
 m_{\nu2} = \sqrt{-\Delta m_{3\ell}^2}\,.
 \end{cases}
\end{equation}
The notation is such that $\Delta m_{3\ell}^2$  is the largest mass splitting eigenvalue
in the case of either mass ordering. $\Delta m_{3\ell}^2 = m_3^2-m_1^2>0$ for a NH and $\Delta m_{3\ell}^2 = m_3^2-m_2^2<0$ for an IH.

\subsubsection{Casas-Ibarra parameterization for \texorpdfstring{$\omega$}{omega}}\label{sec_CasasIbarra}
The parameter $\omega$ can be written as a function of the light neutrino masses, the
PMNS angles and phases, and the heavy Majorana masses $M_p$, using the Casas-Ibarra parameterization~\cite{Casas:2001sr}.
This is a general result whose derivation makes use of the relations in the previous section.
In our case, it reads\footnote{The $i$ factor follows from the conventions chosen above.}
\begin{align}\label{CI_def}
  \omega_{NH} &= \frac{i\sqrt2}{v}\, \diag(\sqrt{M_1}, \sqrt{M_2})\cdot  R \cdot \diag(0, \sqrt{m_{\nu2}}, \sqrt{m_{\nu3}})\cdot  U_{\rm PMNS}^\dag,\\
  \omega_{IH} &= \frac{i\sqrt2}{v}\, \diag(\sqrt{M_1}, \sqrt{M_2})\cdot  R \cdot \diag(0, \sqrt{m_{\nu1}}, \sqrt{m_{\nu2}})\cdot  U_{\rm PMNS}^\dag\label{CI_def_IH}.
  \end{align}
  for the normal and inverted hierarchy respectively. The matrix $R$ is orthogonal and can be parameterized as
\begin{equation}
 R = \begin{pmatrix}
      0& \sqrt{1-r^2}& r\\
      0& -r & \sqrt{1-r^2}
     \end{pmatrix},\qquad r \in \mathbb{C}\,,
\end{equation}
so that the rightmost $2 \times 2$ sub-block is orthogonal. The values of $r$ dictate the relation between measured low energy neutrino parameters and the Lagrangian parameters. Large values of $|r|$ correspond
to parameter space where the $\omega$ and $M$ are related as
\bea\label{R_larger}
\left(\frac{\left(\omega_{2\b}\right)}{\left(\omega_{1\b}\right)}\right)_{|r|\gg1}
\approx \pm i\sqrt{\frac{M_2}{M_1}}.
\eea
The eigenvalues of $c_{\alpha \, \beta}^{(5)}$ result from a significant degree of cancellation between Lagrangian parameters
as a result of this condition being enforced.
Interestingly, the latter emerges naturally if the Majorana states $N_1$, $N_2$ are assumed to form a pseudo-Dirac pair, thereby imposing an approximate  lepton number conservation~\cite{Ibarra:2011xn}.
In the absence of such a symmetry, however, these relations are not invariant under the RGE of the theory, and therefore represent tuned
solutions. To keep the discussion general, in the following we restrict to values $|r| \leq 1$.

\section{Phenomenology of the neutrino option at NLO} \label{sec:NLOanalysis}

Integrating out the $N_p$ states at one loop gives a threshold matching
to the SMEFT Lagrangian parameters proportional to $|\omega|^2/16 \pi^2$ and $|\omega|^4/16 \pi^2$. These threshold matchings
are used to generate the Higgs potential in the neutrino option, so a consistent treatment of the corrections at one loop order is of interest. A more complete treatment of this matching than Ref.~\cite{Brivio:2017dfq} at next to leading order includes all corrections which result from Fig.~\ref{oneloopmatching} and the inclusion of the effects of the running of the Weinberg operator.
It is also necessary to utilize an alternative numerical
strategy than pursued in Ref.\cite{Brivio:2017dfq} to increase the numerical stability of the results. We first develop a consistent NLO framework for studying the neutrino option in this section.
\subsection{One loop matching}
The one loop matching of the seesaw model onto the SMEFT is given by equating
\bea\label{matchingseesaw}
\langle i|\mathcal{L}_{SM} + \mathcal{L}_N|j \rangle = \langle i|\mathcal{L}_{SMEFT} |j \rangle,
\eea
for fixed initial ($i$) and final ($f$) states at the scale $\mu \simeq M$, and solving for the resulting SMEFT parameters. This determines the Wilson coefficients of the higher dimensional operators and defines contributions to the $\mathcal{L}^{(d \leq 4)}$ SM couplings due to matching boundary conditions.

We calculate in dimension regularization with $d=4-2 \epsilon$, and use ${\rm \overline{MS}}$ subtraction. The counterterms renormalize the theories separately on
each side of the matching equations, but finite one loop matching results remain.
\begin{figure}[h!]\centering
 \includegraphics[width=.75\textwidth]{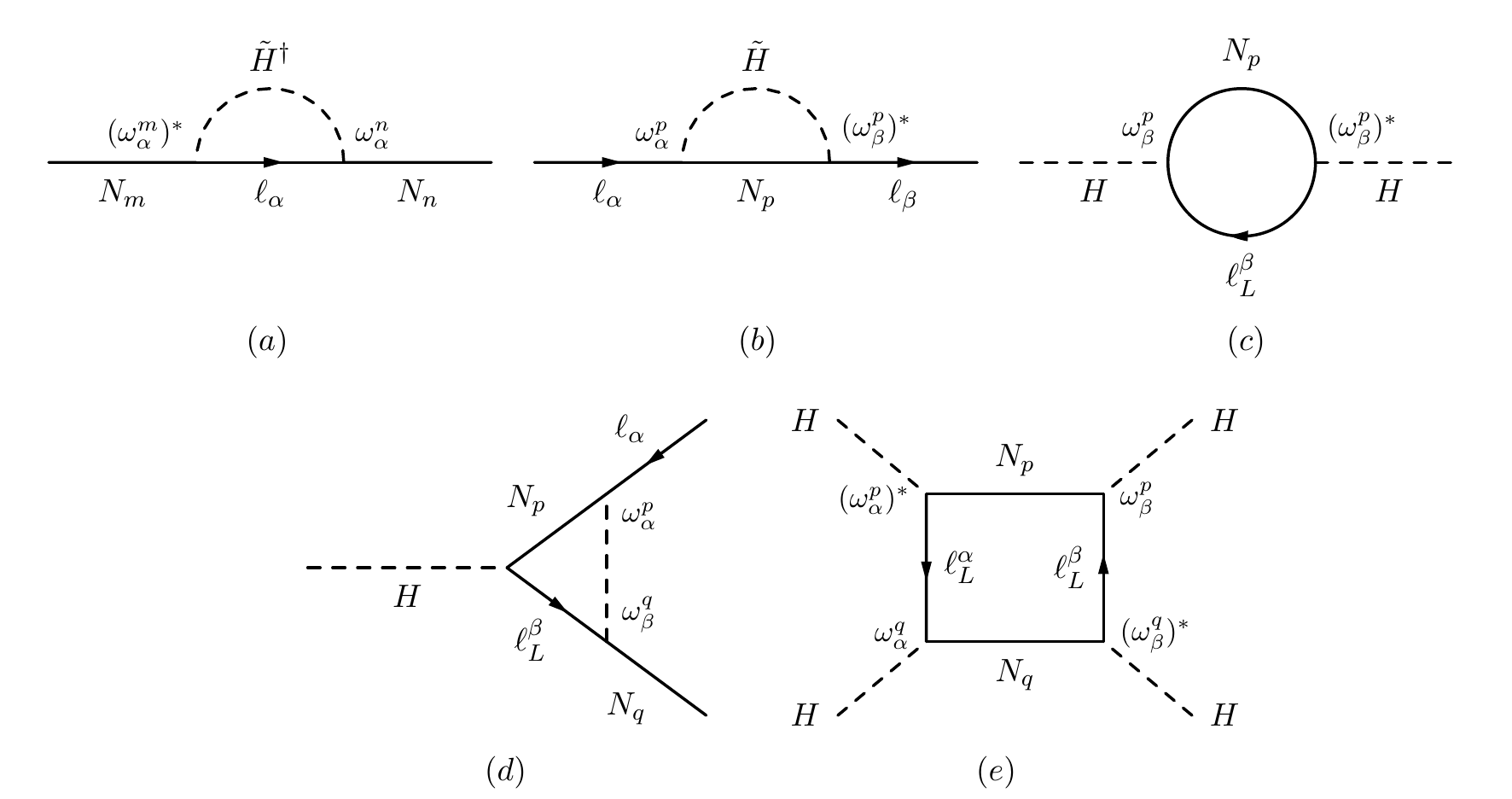}
 \caption{One loop matching diagrams for the seesaw model.}\label{oneloopmatching}
\end{figure}

When matching the propagators used are canonically normalized. The first diagram in Fig.~\ref{oneloopmatching}a leads to a non-canonical $N_p$ field. A canonical normalization condition can be satisfied by performing a finite renormalization field redefinition of the form
\bea
N^{(0)}_{R,m} &\rightarrow \sqrt{R_{mn}^N} \, N^{(r)}_{R,n},
\eea
for the $S$ matrix element in Eqn.\eqref{matchingseesaw}. Here the $(0)$ and $(r)$ superscripts correspond to the un-renormalized and renormalized fields respectively in the on-shell matrix elements
in the seesaw model, and
\bea
R^{N}_{mn} =  \delta_{mn} - \frac{\omega_{\alpha,m} (\omega_n^\alpha)^\dagger}{4 \pi^2} \left(1 + \frac{1}{2}\log \left[\frac{-\mu^2}{M_n^2}\right] \right).
\eea
We have expanded in $m^2/p^2 < 1$ where $m^2$ is the value of the Higgs mass parameter before threshold matching and have fixed $p^2 = M_n^2$ due to the $N$ state which will be taken on-shell
in the threshold matching.
Similarly Fig.~\eqref{oneloopmatching}b leads to a non-canonical kinetic term for $\ell$ which is restored with the finite renormalization field redefinition
\bea
(\ell^{\alpha}_L)^{(0)} \rightarrow   \sqrt{R_{\alpha \beta}^\ell} \,  (\ell^{\beta}_L)^{(r)},
\eea
where
\bea
 R^{\alpha \beta}_\ell  = \delta^{\alpha \beta}
- \frac{\omega_{\alpha,p}^\dagger \omega_\beta^p}{64\pi^2} \left(3 + 2 \log \left[\frac{\mu^2}{M_p^2}\right] \right).
\eea
Similarly the non-canonical kinetic term for $H$ is restored with the finite renormalization field redefinition $
H^{(0)} \rightarrow   \sqrt{R^H} \,  H^{(r)}$
where
\bea
 R^H  = 1- \frac{\omega_{\alpha,p}^\dagger \, \omega_\alpha^p}{32\pi^2} \left(1 + 2 \log \left[\frac{\mu^2}{M_p^2}\right] \right).
\eea

\begin{figure}[h!]\centering
 \includegraphics[width=1\textwidth]{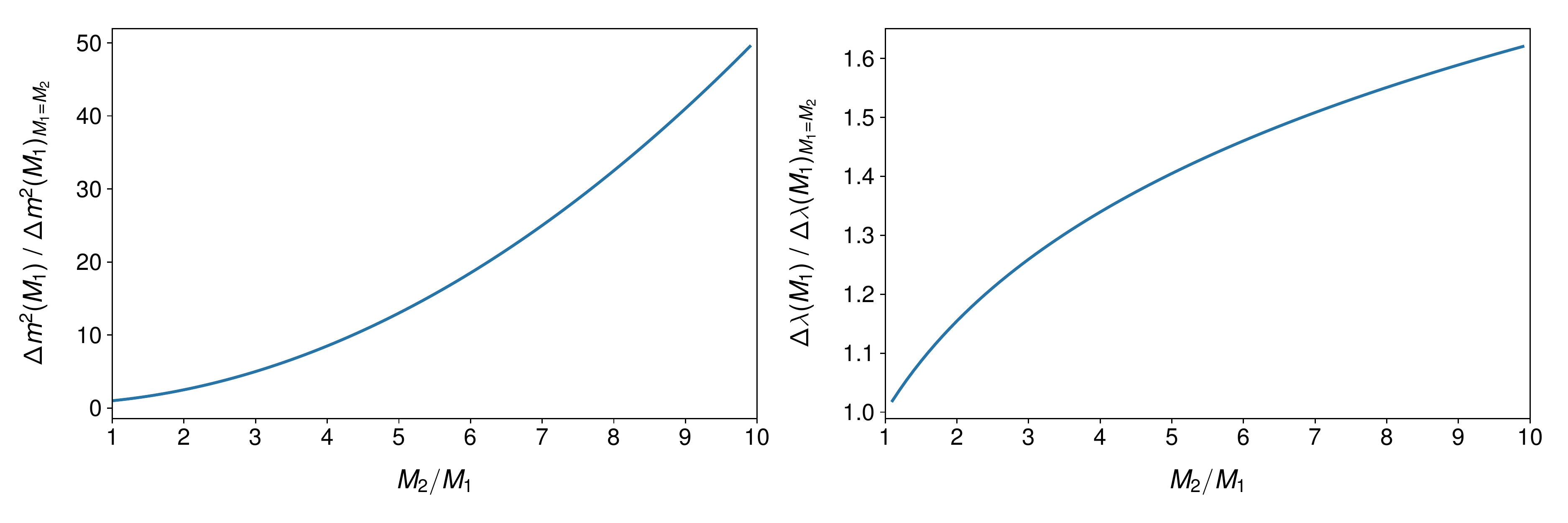}
 \caption{Relative variation of $\Delta \lambda$ (left) and $\sqrt{\Delta m^2}$ (right) assuming $M_2 = x M_1$ compared
 to the degenerate case, as a function of $x$. }\label{plot_thresholds_nondegenerate_variation}
\end{figure}

The three point interactions coupling $N$ to the SM are corrected at one loop. For example, one of the interaction terms has the one loop correction
\bea
- \frac{1}{2} \overline{N_p} \omega_\beta^{p,*}\tilde H ^T \ell_L^{c \, \beta} &\rightarrow&
- \frac{1}{2} \overline{N_p} \omega_\beta^{q,*}\tilde H ^T \ell_L^{c \, \alpha} \left[\sqrt{R^{N, \star}_{qp}}\, \sqrt{R^{\ell, \star}_{\alpha \beta}}  \sqrt{R^{H, \star}} + \frac{\delta_{\alpha \beta}  (\omega_q^\star \cdot \omega_p)}{16 \pi^2} F[\rho_{pq}] \right],
\eea
where
\bea
F[\rho_{pq}] = \int_0^1 \! \! \! dx \int_0^{1-x}  \! \! \! \! \! \! \! \! \! \! dy \frac{2 \sqrt{\rho_{pq}} (y-1)}{x + y \, (x+ y -1) \rho_{pq}},   \quad \rho_{pq} = M_p^2/M_q^2.
\eea
Note that this diagram violates lepton number.
The corrections for the remaining  three point interactions are similar and these (numerically small) effects are required for a complete one loop
treatment of the matching of the seesaw model into the SMEFT.
The threshold corrections to the SM Higgs potential are\footnote{Here we correct an intermediate result in Ref.~\cite{Brivio:2017dfq}. We thank Vedran Brdar for pointing out the $\lambda$ correction.}
\begin{align}
 \Delta \lambda &= \frac{1}{16\pi^2}(\omega_q\cdot\omega^{p*})(\omega_p\cdot\omega^{q*})
 \left(1-\frac{M_pM_q\log\frac{M_p^2}{M_q^2}+M_q^2\log\frac{\mu^2}{M_q^2}-M_p^2\log\frac{\mu^2}{M_p^2}}{M_p^2-M_q^2}\right),\\
 \Delta m^2 &= - \frac{|\omega_p|^2M_p^2}{4\pi^2} \left(1+\log\frac{\mu^2}{M_p^2}\right)\,,
\end{align}
where $\omega_p$ is the $p$-th row of $\omega$.
Restricting to the case of 2 massive right-handed neutrinos: in the degenerate limit $M_1=M_2=M$ and evaluating at $\mu=M e^{-3/4}$ the thresholds
to be consistent with the extraction of the threshold correction from the effective potential in $\rm \overline{MS}$\footnote{See Refs.~\cite{Casas:1998cf,Casas:1999cd} for related results in the effective potential approach.}, the result takes the form
\begin{align}
 \Delta\lambda &= -\frac{5}{32\pi^2}\left(|\omega_1|^4+|\omega_2|^4+6 \abs{\omega_1\omega_2^*}^2 + 4 {\rm Re}(\omega_1 \omega_2)^2 \right),\label{deltalambda_degenerate}\\
 \Delta m^2&=\frac{M^2}{8\pi^2}(|\omega_1|^2+|\omega_2|^2). \label{deltam_degenrate}
\end{align}
In the more general $M_2\gtrsim M_1$ case, we sum the contributions for $p=q=1$ evaluated at $\mu=M_1$, that for $p=q=2$ evaluated at $\mu=M_2$ to be consistent
and the mixed term in $\Delta\lambda$ evaluated at $\mu=M_2$ (where both fields are still dynamical), obtaining:
\bea\label{Dlambda_M1M2}
 \Delta\lambda &=&  -\frac{5}{32\pi^2}\left[|\omega_1|^4+|\omega_2|^4+|\omega_1\omega_2^*|^2\left(1+\frac{2 \, M_1}{M_1 - M_2}\log\frac{M_2^2}{M_1^2}\right) \right], \\
&\, & \, \, + \frac{5}{16\pi^2} \left[ {\rm Re} (\omega_1 \omega_2)^2 \frac{M_1 \, M_2}{M_1^2 - M_2^2} \log \frac{M_1^2}{M_2^2}
 \right], \nn
 \Delta m^2 &=& \frac{1}{8\pi^2}\left[M_1^2|\omega_1|^2+M_2^2|\omega_2|^2\right].\label{Dm2_M1M2}
\eea
Fig.~\ref{plot_thresholds_nondegenerate_variation} shows how the thresholds change assuming $M_2 = x M_1$ relative to the degenerate case, as a function of $x$. For definiteness we have fixed and $\omega_{p\beta}\equiv 1$, but the variation has little dependence on this choice.

The net effect of a consistent one loop matching to the seesaw scenario ($\propto \omega^2/16 \pi^2$) gives the SMEFT at the matching scale
with the Wilson coefficients
\bea
c_{\alpha \, \beta}^{(5)}(\mu^2) &=& \frac{(\omega^p_\alpha)^T \omega_\beta^p}{M_p} \left[1 - \frac{|\omega_p|^2}{4 \pi^2} \left(1 + \frac{1}{2} \log \left[\frac{-\mu^2}{M_p^2}\right] \right) \right]
- \frac{(\omega^p_\alpha)^T \omega_\beta^p}{32 \, \pi^2 \, M_p} \left[ \sum_{s \le p} |\omega_r|^2 \left(1 + 2 \log \left[\frac{\mu^2}{M_r^2}\right] \right) \right], \nn
&-&  \sum_{s \le p} \frac{(\omega^s_\alpha)^T \omega_\beta^p+ (\omega^p_\alpha)^T \omega_\beta^s}{8 \, \pi^2 \, M_p} (\omega^p \cdot \omega_s^\star) \left(1 + \frac{1}{2}\log \left[\frac{-\mu^2}{M_s^2}\right]
- \frac{F[\rho_{sp}]}{4}  \right), \\
&-& \frac{(\omega^s_\alpha)^T \omega_\beta^p+ (\omega^p_\alpha)^T \omega_\beta^s}{8 \, \pi^2 \, M_p} \frac{\epsilon_{sp} (M_1+M_2)}{M_1-M_2} \Re(\omega_1 \cdot \omega_2^\star) \left[1 + \frac{1}{2} \log \left[\frac{-\mu^2}{M_2^2}\right]\right],  \nn
&-& \frac{(\omega^p_\rho)^T \omega_\beta^p}{16 \, \pi^2 \, M_p}  \sum_{s \le p} \frac{\omega_\alpha^s (\omega_\rho^s)^\dagger}{8} \left(3 + 2 \log \left[\frac{\mu^2}{M_s^2}\right] \right)
- \frac{(\omega^p_\alpha)^T \omega_\sigma^p}{16 \, \pi^2 \, M_p}  \sum_{s \le p} \frac{\omega_\beta^s (\omega_\sigma^s)^\dagger}{8} \left(3 + 2 \log \left[\frac{\mu^2}{M_s^2}\right] \right), \nonumber
\eea
with $s$ summed over the $N_s$ active states, when each $N_p$ mass eigenstate Majorana field is integrated out. $\epsilon_{12} = - \epsilon_{21} = 1$ and $M_{1,2}$ are the tree level masses of the $N^p$ states.
The potential after the one loop matching is given by
\bea
 V(H^\dag H) &=& -\frac{m_0^2  \, R_H + \Delta m^2}{2}(H^\dag H) + (\lambda_0 \,  R_H^2 + \Delta \lambda) (H^\dag H)^2 + \cdots.
\eea
We do not assume that hierarchies among the $\omega^p_\beta$, or significant effects due to their orientation in flavour space  in order to enhance the importance of these one loop corrections.
Further, we assume $m_0$ is negligible as we are considering a classically scaleless mass spectrum. In the numerical results presented we absorb
the correction to $\lambda_0, M_p$ into the leading order parameters as a result and neglect the correction to $c_{\alpha \, \beta}^{(5)}$.\footnote{The one loop results given here can be compared to some overlap with past results in the literature given in
Refs.~\cite{Grimus:1989pu,Grimus:2002nk,Dev:2012sg,Fernandez-Martinez:2015hxa}.}
We adopt this approach as these corrections are  smaller than the remaining numerical uncertainties in the RGE evolution of the SMEFT parameters.
\subsection{Running}
The SM potential parameters are compared to measured values run to the
scale $\mu=\hat{m}_t$.
For the RGE of the SM we use results from Appendix B of Ref.~\cite{Buttazzo:2013uya} and evaluate the running of $\{g_1, g_2, g_2, y_t, y_b, y_\tau, \lambda, m^2\}$
as a coupled system, using the RGE computed at $n_{RGE}$ loops. The results in Ref.~\cite{Buttazzo:2013uya} allow numerical studies
of the order $n_{RGE}=3$ for $\{g_1, g_2, g_2, y_t, \lambda, m^2\}$ (without $y_b,\,y_\tau$ dependence) and up to $n_{RGE}=2$ for $\{y_b,y_\tau\}$.

\subsection{Neutrino parameter running}
The seesaw model matched onto $c_{\alpha \, \beta}^{(5)}(\mu^2 = M^2)$ is compared to the
values of masses and mixing angles extracted at $\mu = \hat{m}_Z$ after running the seesaw parameters
with one loop RGE evolution of the Weinberg operator.\footnote{Here a hat superscript indicates a experimentally measured quantity.}
The RGEs of  the $c_5$ coefficient are extracted from Refs.~\cite{Babu:1993qv,Antusch:2001ck}.
Below the scale $\hat{m}_Z$ the neutrino parameters do not run significantly, as
we have explicitly verified.

The RGE of the $c_5$ coefficient is \cite{Babu:1993qv,Antusch:2001ck}:
\begin{equation}\label{running_c5}
\begin{aligned}
 16\pi^2\mu\frac{d c_{5}}{d\mu} &= -\frac{3}{2} \left[c_5 (Y_e^\dag Y_e) + (Y_e^\dag Y_e)^T c_5\right] - \left[3 g_2^2- 4\lambda-2\left(3Y_u^\dag Y_u+3Y_d^\dag Y_d+Y_e^\dag Y_e\right)\right]\, c_5\\
 &\simeq -\frac{3}{2} \left[c_5 \cdot\diag(0,0,y_\tau^2)+\diag(0,0,y_\tau^2)\cdot c_5\right] - \left[3 g_2^2- 4\lambda-6y_t^2-6y_b^2-2y_\tau^2\right]\, c_5\,.
 \end{aligned}
\end{equation}
The Yukawa coupling normalization is  $\mathcal{L}_Y = -\biggl[ H^{\dagger j} \overline d\, Y_d\, q_{j}
+ \widetilde H^{\dagger j} \overline u\, Y_u\, q_{j} + H^{\dagger j} \overline e\, Y_e\,  \ell_{j} + \hbox{h.c.}\biggr]$.
In this limit, there is no mixing between the different entries of $c_5$ and the individual elements run according to
\begin{equation}
 16\pi^2\mu\frac{d c_{5,\a\b}}{d\mu} = -\kappa_{\a\b} \,c_{5,\a\b},\qquad \kappa_{\a\b}=\begin{cases}
                                                                                      3 g_2^2- 4\lambda-6y_t^2-6y_b^2-2y_\tau^2, & \a,\b\neq 3\\
                                                                                      3 g_2^2- 4\lambda-6y_t^2-6y_b^2-y_\tau^2 / 2, & \b(\a)=3,\, \a(\b)\neq 3\\
                                                                                      3 g_2^2- 4\lambda-6y_t^2-6y_b^2+y_\tau^2, & \a=\b=3\,.
                                                                                     \end{cases}
\end{equation}

Comparing the extracted eigensystem of $c_5$
at any given scale $\mu$ to experimental results in parameter scans is numerically unstable.
Using the $\b$-functions of the measurable neutrino parameters themselves reduces this numerical uncertainty. We use the results in Ref.~\cite{Casas:1999tg} for a generic parameterization\footnote{Corrected
by a factor of three reported in Ref.~\cite{Antusch:2001ck}.} of the measurable parameters extracted from $c_5$. For a normal hierarchy\footnote{These expressions have been derived from the general parameterization in Ref.~\cite{Casas:1999tg} imposing $m_{\nu1}=0$. The case for inverted hierarchy can be inferred analogously, choosing $m_{\nu3}=0$.} with two nonzero masses, the results in Ref.~\cite{Casas:1999tg}
reduce to
\begin{align}\label{beta_nuPar_1st}
 16\pi^2\mu \frac{d m_{\nu,k}}{d\mu} &= - m_{\nu, k} \left[3 y_\tau^2 |V_{3i}|^2 + 3 g_2^2- 4\lambda-6y_t^2-6y_b^2-2y_\tau^2\right]\\
 16\pi^2\mu \frac{d \theta_1}{d\mu} &= \frac{3y_\tau^2}{2c_2} \,\Re\left[
 s_3 V_{31}V_{33}^*+ c_3 \, T_{32} \right]\\
 16\pi^2\mu \frac{d \theta_2}{d\mu} &= \frac{3y_\tau^2}{2}\,\Re\left[e^{-i \delta}\left(
 -c_3 V_{31}V_{33}^*+ s_3 \, T_{32}
\right)\right]\\
 16\pi^2\mu \frac{d \theta_3}{d\mu} &= -\frac{3y_\tau^2}{2}\frac{s_2}{c_2}\,\Re\Bigg[
 e^{-i \delta}\Bigg(
\frac{c_2}{s_2}e^{i\delta} V_{31}V_{32}^*+s_3 V_{31} V_{33}^*
- c_3 \, T_{32}
\Bigg)\Bigg]
\end{align}
Here we have used the notation
\bea
T_{32} = \frac{2m_{\nu2}m_{\nu3}}{m_{\nu2}^2-m_{\nu3}^2} e^{i\phi'} V_{32}^* V_{33}
 + \frac{ m_{\nu2}^2+m_{\nu3}^2}{m_{\nu2}^2-m_{\nu3}^2}V_{32} V_{33}^*.
\eea
The remaining RGEs are
\begin{align}
16\pi^2\mu\frac{d \delta}{d\mu} &=\frac{3y_\tau^2}{2}\, \Im\Bigg[
\frac{V_{31}V_{32}^*}{c_3s_3}
-\frac{s_3}{s_1c_2c_3}V_{31}V_{22}^*V_{33}^*
- \frac{e^{-i \delta}}{s_2 c_1 c_2}V_{31}V_{22}V_{33}^*
-T_{32} \left(\frac{e^{- i \delta} V_{21}}{c_1 \, c_2 \, s_2} - \frac{c_3 V_{21}^\star}{s_1 s_3 c_2} \right)
\Bigg]\\
16\pi^2\mu \frac{d \phi}{d\mu} &=3y_\tau^2 \, \Im\Bigg[
\frac{c_3 V_{31}V_{32}^*}{s_3} + \frac{V_{21}^*V_{33}^* V_{31}}{s_1 c_2}
+\frac{|V_{31}|^2}{c_1c_2}V_{33}^* +  T_{32}
\left(\frac{c_3 V_{21}^*}{s_1c_2s_3}-\frac{V_{32}^*}{c_1c_2}\right)
\Bigg]\\
16\pi^2\mu  \frac{d \phi'}{d\mu} &=3y_\tau^2\, \Im\Bigg[
\frac{s_3 V_{31}V_{32}^*}{c_3} +\frac{|V_{31}|^2V_{33}^*}{c_1c_2}
-\frac{s_3}{s_1c_2c_3}V_{31}V_{22}^*V_{33}^* - T_{32}\left(\frac{V_{22}^*}{s_1c_2}+\frac{V_{32}^*}{c_1c_2}\right)
\Bigg]
\end{align}
where $V_{ij}$ denotes the corresponding entry of the $V$ matrix defined in Eq.~\eqref{PMNS_def}.

Ref.~\cite{Casas:1999tg} defined the running of three unphysical phases, that are added to the definition of the $U$ rotation such that
\begin{equation}
 U = \diag\left(e^{i\a_e},e^{i\a_\mu},e^{i\a_\tau}\right)\cdot U_{\rm PMNS}\,.
\end{equation}
This approach is convenient as the field redefinitions that reduce the phases of the PMNS matrix to the minimal set must be re-imposed at each scale $\mu$.
The $\beta$ functions of the unphysical phases are
\begin{align}
 16\pi^2\mu  \frac{d \a_e}{d\mu} &= \frac{3y_\tau^2}{2}
 \Im \Bigg[\frac{V_{31} V_{32}^*}{s_3 c_3} + \frac{|V_{31}|^2 V_{33}^*}{c_1 c_2} - \frac{s_3}{s_1 c_2 c_3} V_{31} V_{22}^* V_{33}^*
 - T_{32} \left(\frac{V_{32}^*}{c_1c_2}+\frac{V_{22}^* c_3}{s_1 c_2 s_3}\right)
 \Bigg], \\
16\pi^2\mu  \frac{d \a_\mu}{d\mu} &= \frac{3y_\tau^2}{2c_2s_1}\, \Im\Bigg[
V_{31}V_{21}^*V_{33}^* - T_{32} V_{22}^*
\Bigg], \\
16\pi^2\mu \frac{d \a_\tau}{d\mu} &=\frac{3y_\tau^2}{2c_2c_1}\, \Im\Bigg[
|V_{31}|^2V_{33}^*  - T_{32} V_{32}^*
\Bigg].\label{beta_nuPar_last}
\end{align}
We have verified analytically that
\begin{equation}
 \left(16\pi^2\mu\frac{d c_{5}}{d\mu}\right) = \sum_k \frac{\de c_{5}}{\de x_k} \left(16\pi^2\mu\frac{d x_k}{d\mu}\right),\qquad x_k=\{m_{\nu1},m_{\nu2},m_{\nu3},\theta_1,\theta_2,\theta_3,\delta,\phi,\phi',\a_e,\a_\mu,\a_\tau\}\,.
\end{equation}
The running of neutrino parameters depends on $\{g_2,\lambda,y_t,y_b,y_\tau\}$.
In the numerical results, these are fixed to their value at $\mu=m_t$ neglecting yet higher order running effects.
Although the running of the neutrino parameters was studied for completeness in the numerical results shown, we confirm past results that
find the neutrino parameters do not run in a numerically significant manner. We have verified that the dependence on the top quark mass in the
numerical running of the neutrino parameters is a negligible effect when scanning parameter space.

\section{Numerical strategy and results} \label{sec:numerics}
Consistency of the neutrino option explaining the Higgs potential and Neutrino masses is dictated by choosing a subset of
inputs among
\bea
\{\hat{m}_\nu^i(\mu^2 \ll \langle H^\dagger H \rangle), \hat{\theta_i}, \hat{m}_h(\langle H^\dagger H \rangle), \hat{\lambda}(\langle H^\dagger H \rangle)  \},
\eea
and predicting the remaining quantity(ies). In Ref.~\cite{Brivio:2017dfq} a consistency test was formulated where $\{\hat{\lambda}(M), \hat{m}_h(M)\}$ was fixed,
and it was shown that parameters can be chosen such that $\hat{m}_\nu^i(\mu^2 \ll \langle H^\dagger H \rangle)$ can be approximately reproduced.
It was observed that significant numerical instability is present in this approach.
It is necessary to avoid an asymptotic value of $\lambda(M) \rightarrow 0$ being used as an input for numerical precision.
In a consistency test, any mismatch between a predicted value of a parameter, and an observed value can be accommodated by an extended scenario where the Neutrino
option is embedded in a UV theory. Only the parameter $\lambda$ can receive further classical tree level contributions without breaking a symmetry
or adding field content to the scenario. For these reasons, in this paper we formulate a consistency test
where $\{\hat{m}_\nu^i(\mu^2 \ll \langle H^\dagger H \rangle), \hat{\theta_i}, \hat{m}_h(\langle H^\dagger H \rangle)\}$ are fixed as inputs, and the required $\lambda(M)$ is then compared
to $\lambda_0(M) + \Delta \lambda(M)$.

\subsection{Numerical inputs}
We enforce that the low energy neutrino parameters taken from the global fit in Ref.~\cite{Esteban:2016qun}
\begin{table}[h!]\centering
\renewcommand{\arraystretch}{1.3}
\begin{tabular}{>{$}l<{$}|cc|cc}
\hline
& \multicolumn{2}{c|}{Normal Hierarchy}& \multicolumn{2}{c}{Inverted Hierarchy}\\
& best fit& $3\s$ range&  best fit& $3\s$ range\\\hline
s_1^2& 0.441 & 0.385 -- 0.635 & 0.587& 0.393 -- 0.640\\
s_2^2& 0.02166& 0.01934 -- 0.02392& 0.02179 & 0.01953 -- 0.02408\\
s_3^2& 0.306& 0.271 --0.345 & 0.306 & 0.271 -- 0.345\\
\d (^\circ) & 261& 0 -- 360 & 277 & 145 -- 391\\
\Delta m_{21}^2\, ( \unit[10^{-5}]{eV^2})& 7.50 & 7.03 -- 8.09 & 7.50& 7.03 -- 8.09\\
\Delta m_{3l}^2\,\, ( \unit[10^{-3}]{eV^2})& 2.524 & 2.407 -- 2.643& -2.514& (-2.635) -- (-2.399)\\
\hline
\end{tabular}
 \caption{Best fit values of neutrino parameters taken from the global fit
 in Ref.~\cite{Esteban:2016qun}.\label{latest_nu_fit_intervals}}
\end{table}
and given in Table.~\ref{latest_nu_fit_intervals} are reproduced.

The SM parameters are extracted from Ref.~\cite{Buttazzo:2013uya} solving
the reported integral equations for the input parameters $\hat{m}_W, \hat{m}_Z, \hat{m}_h$
at the indicated loop order. The light bottom and $\tau$ Yukawa's,
$\hat{y}_b,\hat{y}_\tau$, are matched at tree level as higher order corrections are negligible. The value of $g_3(\mu=m_t)$ is extracted from  Eqn.~60 of Ref.~\cite{Buttazzo:2013uya} which includes
higher order QCD corrections.

To these SM results we add the effects of the seesaw model as matched onto the SMEFT. The results in Ref.~\cite{Elgaard-Clausen:2017xkq} characterize the tree level matching of the seesaw model onto the SMEFT
up to sub-leading order ($\mathcal{L}^{(7)}$ corrections) but we restrict our attention to the matching onto $\mathcal{L}^{(5)}$ in this work.

\begin{table}[h!]\centering
\begin{tabular}{lc@{\hspace*{-2mm}}c}
 \hline
& best fit& range\\\hline
$\hat{G}_F$ [GeV$^{-2}$]       & 1.1663787 $\cdot 10^{-5}$&\\
$\hat{\a}_s(m_Z)$ & 0.1185&\\
$\hat{m}_Z $ [GeV]     & 91.1875&\\
$\hat{m}_W $ [GeV]     & 80.387&\\
$\hat{m}_h $ [GeV]     & 125.09& \\
$\hat{m}_t $ [GeV]     & 173.2&171 -- 175\\
$\hat{m}_b $ [GeV]     & 4.18&\\
$\hat{m}_\tau$ [GeV]   & 1.776&\\\hline
\end{tabular}
~~~
\begin{tabular}{lccc}
\hline
& tree& 1-loop&  2-loop\\    \hline
$ \hat{\lambda}$&       0.1291&	  0.1276&	0.1258\\
$ \hat{m}$ [GeV]&            125.09&         132.288&      131.431\\
$ \hat{g}_1$&           0.451&	  0.463&	0.461\\
$ \hat{g}_2$&           0.653&	  0.6435&	0.644\\
$ \hat{g}_3$&           \multicolumn{3}{c}{------ 1.22029 ------}\\
$ \hat{y}_t$&           0.995&	  0.946&	0.933\\
$ \hat{y}_b$&           0.024&	  -&	-\\
$ \hat{y}_\tau$&        0.0102&	  -&     - \\\hline
\end{tabular}
\caption{Left table: best fit values of the quantities used as inputs in the numerical analysis, while $m_t$ is varied in the range specified. Right table: matching values for the SM parameters at $\mu=m_t$ obtained from the expressions in Appendix A in Ref.~\cite{Buttazzo:2013uya} with the inputs on the left
when $m_t=173.2 \, {\rm GeV}$.\label{tab_SM_matching_values}}
\end{table}

The $\omega_{p\b}$ are required to reproduce the observed neutrino masses and mixings. The range of values for $\Delta m^2(M_1),\, \Delta\lambda(M_1)$ compatible with this condition is determined
scanning the low energy parameter space with a sample of 1000 points randomly selected within the $3\sigma$ allowed ranges for $\{\hat{m}_{\nu i}, \hat{\theta}_i, \hat{\delta}, \hat{\phi},\hat{\phi'}\}$ given in Table~\ref{latest_nu_fit_intervals}.

Each point determined represents a boundary condition at $\mu=\hat m_Z$ for the neutrino parameters' RGE (Eqns.~\eqref{beta_nuPar_1st} - \eqref{beta_nuPar_last}). For each of them it is then possible to determine the running quantities $\omega_{p\beta}(\mu)$ via the Casas-Ibarra parameterization (Eqns.~\eqref{CI_def}, \eqref{CI_def_IH}) and consequently the threshold corrections $\Delta m^2(\mu=M_1)$, $\Delta\lambda(\mu=M_1)$ as a function of the lightest Majorana mass $M_1$ (Eqns.~\eqref{Dm2_M1M2}, \eqref{Dlambda_M1M2}).
The parameter $r$ of the Casas-Ibarra parameterization is varied at every point and chosen as a random complex with $|r|\leq1$. Four independent scans are performed; assuming either normal or inverted neutrino mass hierarchy and either degenerate ($M_1=M_2$) or nearly-degenerate ($M_1\lesssim M_2\lesssim 10\, M_1$) $N_p$ states.\footnote{We do not consider cases where $M_2\gg M_1$ as a different numerical treatment would be required in this case. The choice of nearly-degenerate Majorana states can be consistent with resonant leptogenesis \cite{Pilaftsis:2003gt}.}

The values of $\{m^2(\mu),\lambda(\mu)\}$ that are compatible with the measured $\{\hat m_h$, $\hat\lambda=\hat{G}_F \hat m_h^2/2\}$ are then determined.
These are the solutions to the SM RGE system~\cite{Buttazzo:2013uya} with the matching conditions in Table~\ref{tab_SM_matching_values} (right) fixed at $\mu=\hat m_t$.
We consider RGEs with $n_{RGE}=\{1,\,2,\,3\}$ with order ($n_{RGE}-1$) matching and three benchmark values for $\hat m_t=\{171,\,173.2,\,175\}$~GeV.

We then compare the results obtained in these steps. Unlike in Ref.~\cite{Brivio:2017dfq}, for the sake of generality we allow here for a term $\lambda_0 (H^\dag H)^2$ in the scalar potential.
The neutrino option is then realized for values of ($M_1$, $\lambda_0$) that satisfy simultaneously
\begin{subequations}\label{NO_conditions}
\begin{align}
m^2(M_1) &\simeq \Delta m^2(M_1),\label{NO_conditions_m2}\\
\lambda(M_1) &\simeq \lambda_0 + \Delta \lambda(M_1)\,. \label{NO_conditions_lam}
\end{align}
\end{subequations}
\subsection{Case \texorpdfstring{$m_t=173.2$~GeV}{mt=173.2 GeV} and normal neutrino mass hierarchy}
\begin{figure}[t]\centering
\includegraphics[width=.6\textwidth]{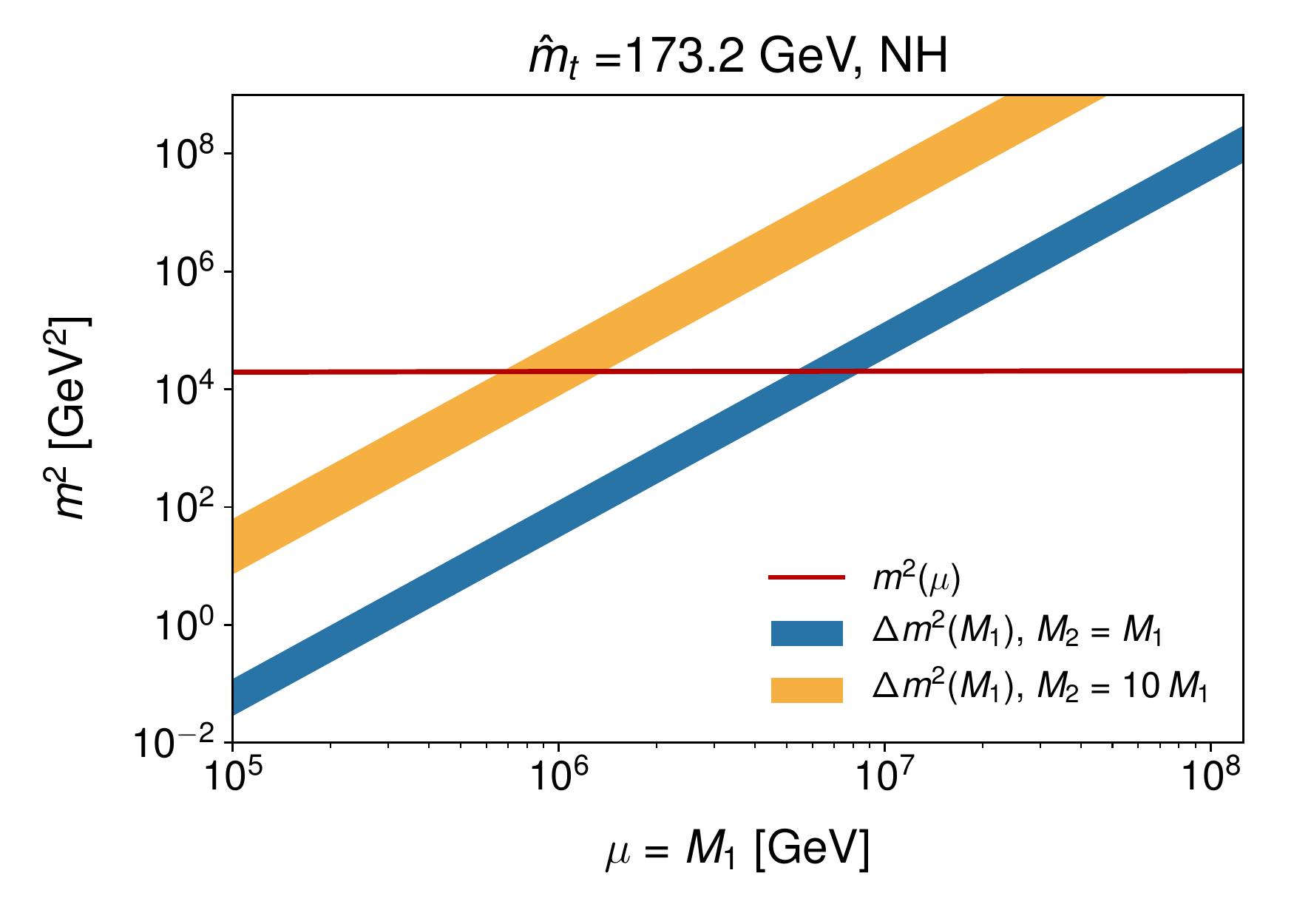}
\caption{Numerical comparison between the values of the threshold correction $\Delta m^2(M_1)$ compatible with neutrino physics constraints in the degenerate $M_1=M_2$ case (blue band) and when $M_2 = 10\,M_1$ (orange band) with the running Higgs mass $m^2(\mu)$ determined by the SM RGE and the measured SM parameters (red line). The running effect is not appreciable in the scale shown because $m^2(\mu)$ varies vary little compared to $\Delta m^2(M_1)$. This plot assumes normal ordering of neutrino masses (NH) and $\hat m_t = 173.2$~GeV. }\label{plot.num_matching_m2}
\end{figure}
\begin{figure}[t]\centering
\hspace*{-15mm}
\includegraphics[width=.553\textwidth]{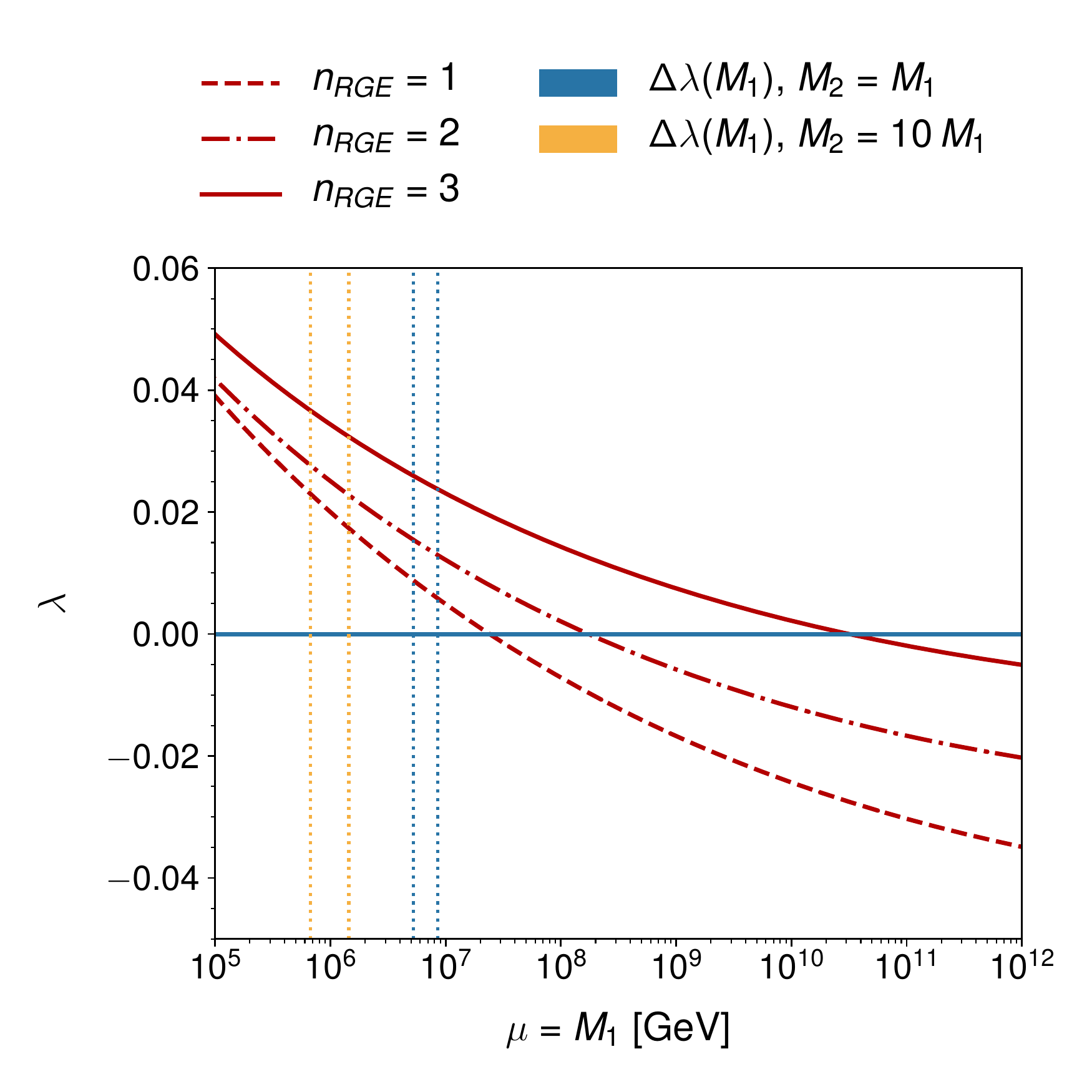}
\hspace*{-5mm}
\includegraphics[width=.553\textwidth]{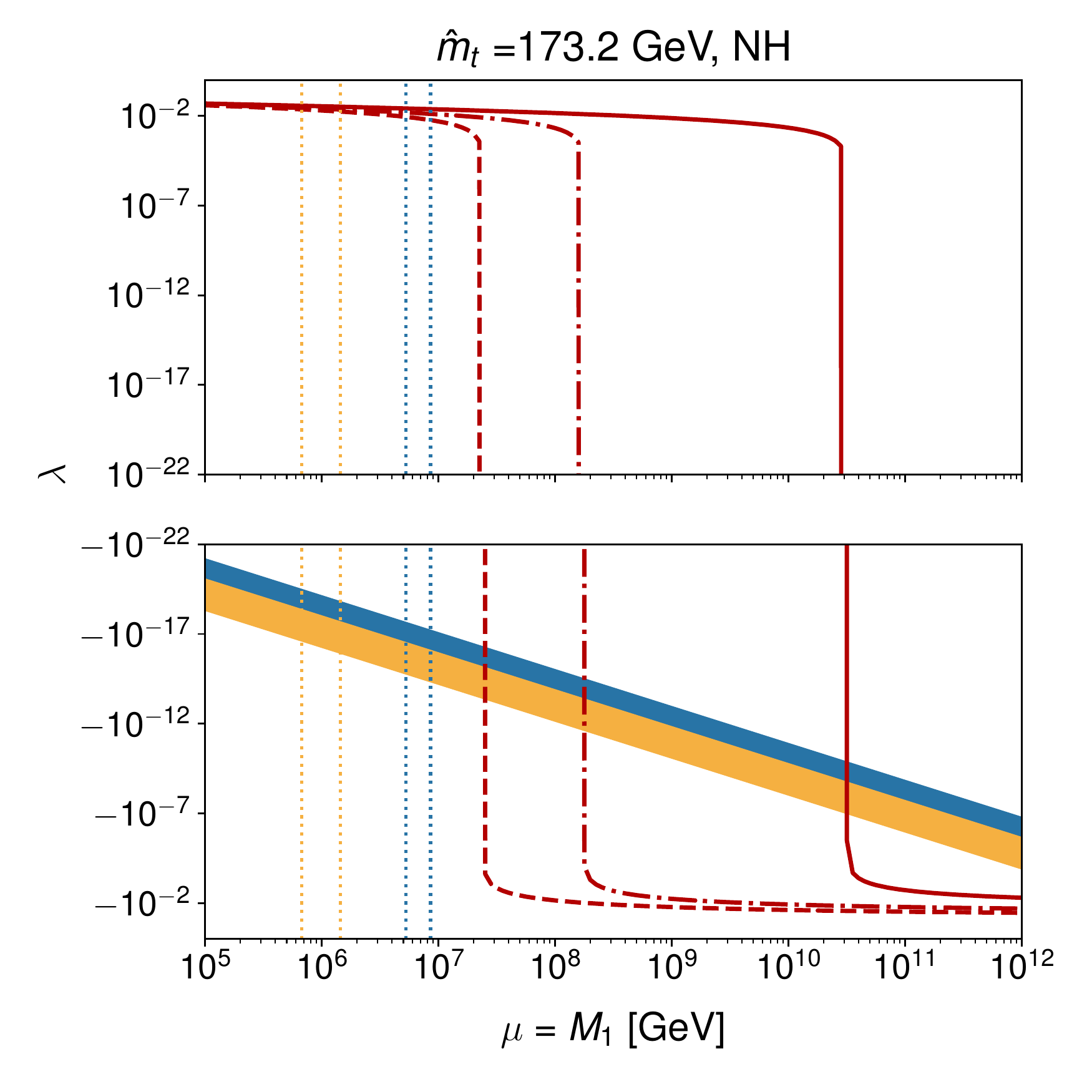}
\caption{Numerical comparison between the values of the threshold correction $\Delta \lambda(M_1)$ compatible with neutrino physics constraints in the degenerate $M_1=M_2$ case (blue band) and when $M_2 = 10\,M_1$ (orange band) with the running quartic coupling $\lambda(\mu)$ determined by the SM RGE and the measured SM parameters (red lines). The blue and orange bands overlap completely in the left plot.
The left and right plots differ uniquely in the choice of the $y$-axis scale (linear vs. logarithmic) and the right figure has been split in two symmetric panels for $\lambda>0$ and $\lambda<0$.
 The dotted vertical lines mark the mass regions where the matching for $m^2$ is fulfilled (cf. Fig.~\ref{plot.num_matching_m2}) for both the degenerate (blue) and non-degenerate (orange) cases.
This figure assumes normal ordering of neutrino masses (NH) and $\hat m_t = 173.2$~GeV.
 }\label{plot.num_matching_lam_173_NH}
\end{figure}

The results of the analysis are shown in Figs.~\ref{plot.num_matching_m2},~\ref{plot.num_matching_lam_173_NH} for the case of normal neutrino mass hierarchy and $\hat m_t = \unit[173.2]{GeV}$.
Fig.~\ref{plot.num_matching_m2} shows $m^2(\mu)$ (red line) vs. $\Delta m^2(M_1)$ for degenerate $N_p$ states (blue band) and for $M_2= 10\, M_1$ (orange band). Eqn.~\eqref{NO_conditions_m2} is satisfied in the region where the bands overlap with the RGE curve: for the degenerate case we find\footnote{Note that numerical subtleties of scanning parameter space using the Casas-Ibarra parameterization are known, see Ref.~\cite{Casas:2010wm} for a discussion. Exceptional parameter space can possibly exist outside the results shown, which are inferred from the numerical procedure above. The shown regions are expected to determine the bulk of the allowed parameter space.}
\begin{equation}\label{M_range_result}
\begin{aligned}
5 \cdot\unit[10^6]{GeV}
\;\lesssim\; &\;M_1 \;\lesssim\;
8.3 \cdot\unit[10^6]{GeV}&  &({\rm NH})\,,
\\
4.2 \cdot\unit[10^6]{GeV}
\;\lesssim\; &\;M_1 \;\lesssim\;
7 \cdot\unit[10^6]{GeV}&  &({\rm IH})\,,
\end{aligned}
\end{equation}
while for $M_2/M_1 = x >1$ lower values of $M_1$ are allowed. For the benchmark $x=10$ the viable mass region is
\begin{equation}\label{M_range_result_nondeg}
\begin{aligned}
6.4 \cdot\unit[10^5]{GeV}
\;\lesssim\; &\;M_1 \;\lesssim\;
1.4 \cdot\unit[10^6]{GeV}&  &({\rm NH})\,,
\\
5.5 \cdot\unit[10^5]{GeV}
\;\lesssim\; &\;M_1 \;\lesssim\;
8.9 \cdot\unit[10^5]{GeV}&  &({\rm IH})\,,
\end{aligned}
\end{equation}
and intermediate values are possible for $1<x<10$.
Notably, $m^2(\mu)$ has negligible dependence on both $n_{RGE}$ and $\hat m_t$. Therefore the range
\begin{equation}\label{M_range_result_general}
5\cdot \unit[10^5]{GeV}
\;\lesssim\; M_1 \;\lesssim\;
\unit[10^7]{GeV}\,
\end{equation}
is a general prediction of the neutrino option.
Specific assumptions about the neutrino mass ordering and $M_2/M_1$ refine this range as detailed in Figure~\ref{plot.logm1_lam}.

\begin{figure}[t]\centering
\hspace*{-15mm}
\includegraphics[width=.553\textwidth]{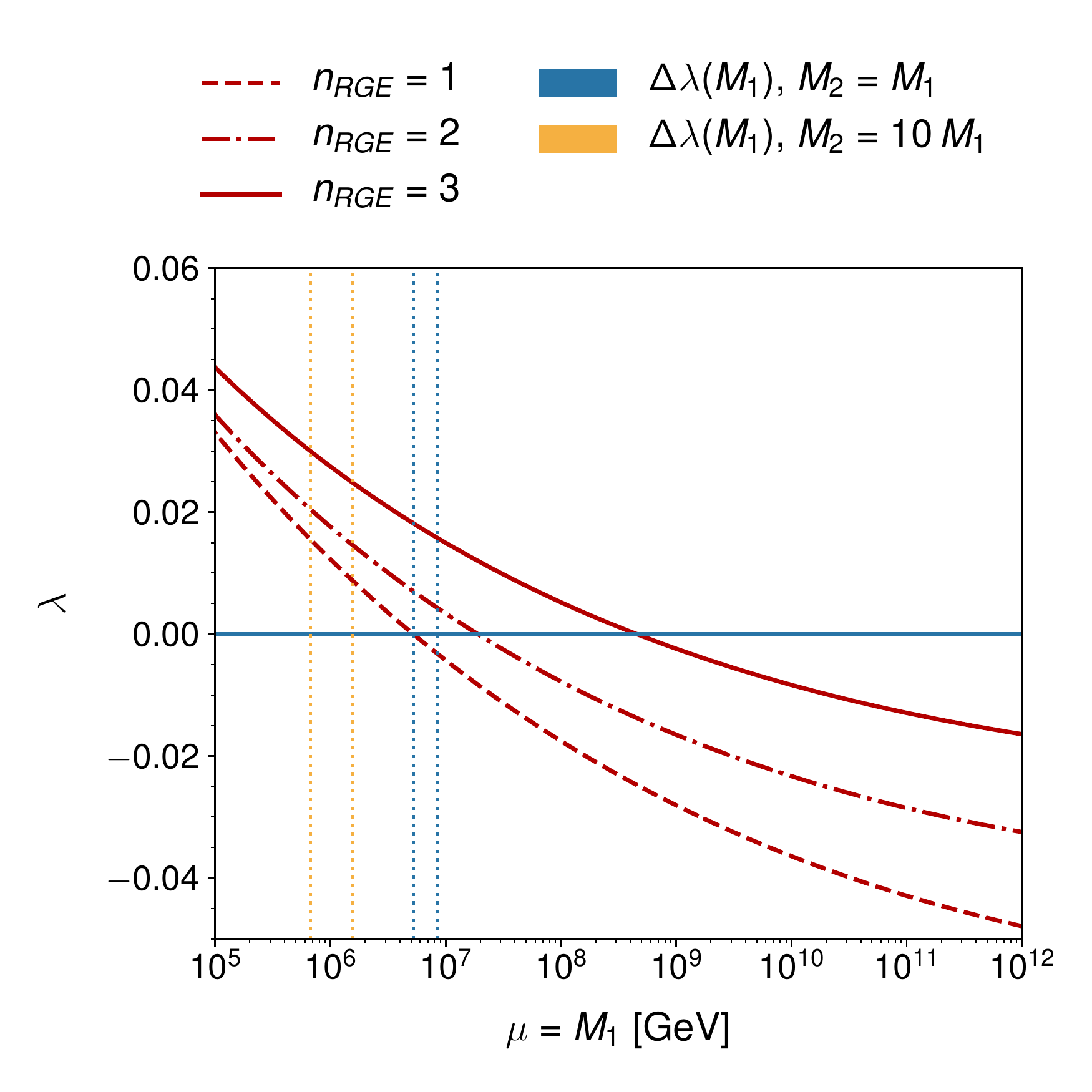}
\hspace*{-5mm}
\includegraphics[width=.553\textwidth]{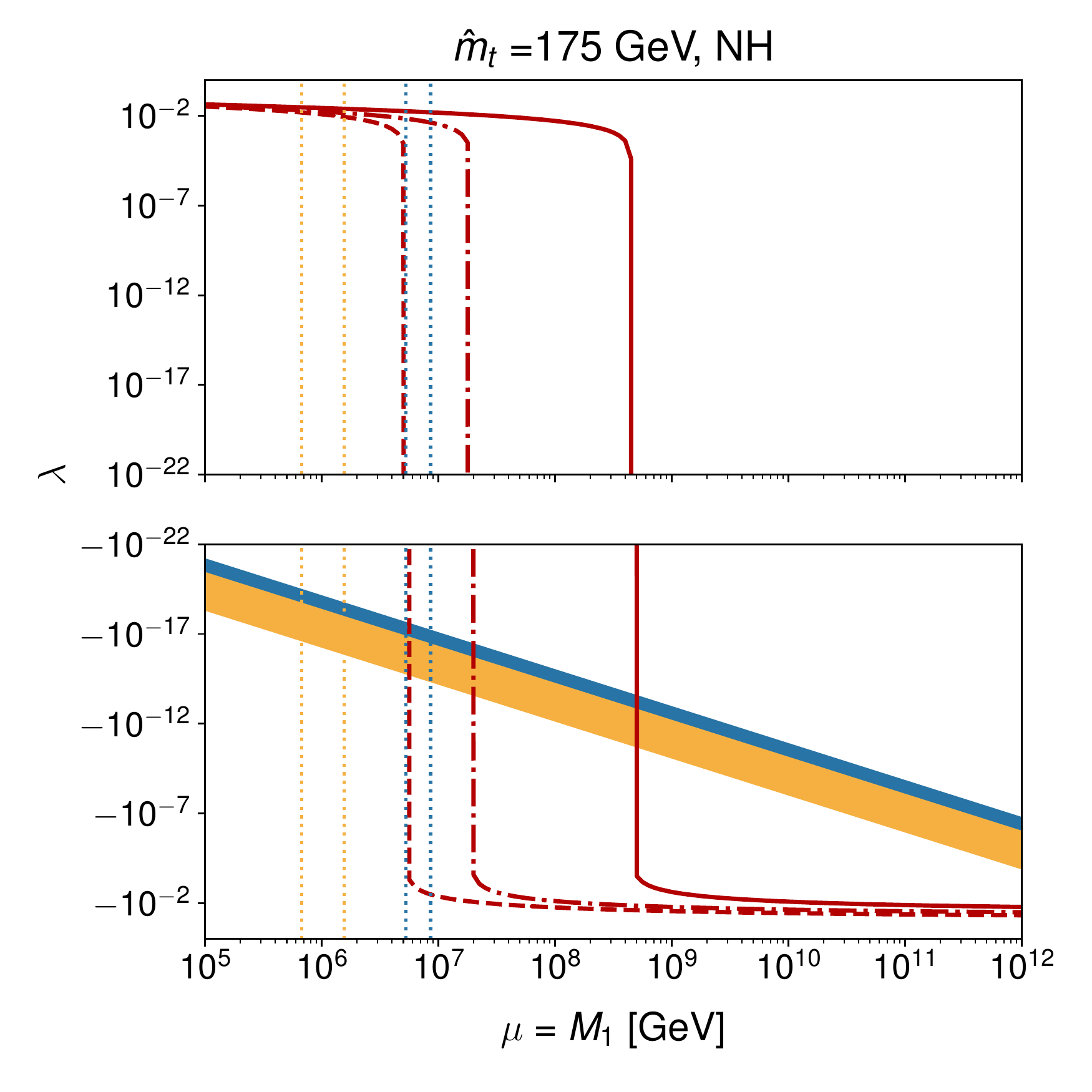}
\caption{Same as Fig.~\ref{plot.num_matching_lam_173_NH}, for a top mass $\hat m_t=\unit[175]{GeV}$.}\label{plot.num_matching_lam_NH_175}
\end{figure}
Fig.~\ref{plot.num_matching_lam_173_NH}  shows $\lambda(\mu)$  vs $\Delta\lambda(M_1)$, both in linear (left) and log scale (right). For reference, the regions in Eqns.~\eqref{M_range_result},~\eqref{M_range_result_nondeg} are  marked with blue and orange dotted vertical lines. Within these energy windows  the threshold correction $\Delta\lambda(M_1)$ is always negative and very small, while the SM running curve $\lambda(\mu)$ is positive and $\sim\mathcal{O}( 10^{-2})$. For the neutrino option to be realized it is then necessary that
\begin{equation}
\lambda(M_1)\simeq\lambda_0 \gg \Delta\lambda(M_1),\quad\Rightarrow\quad \lambda_0 \sim\mathcal{O}(10^{-2})\,.
\end{equation}

\subsection{Varying \texorpdfstring{$m_t$}{mt} and other benchmark assumptions}
Assuming an inverted neutrino mass hierarchy and taking different values of the top quark mass leads to qualitatively similar figures and has a modest impact on the numerical results.
The main conclusions, i.e. the identification of the mass range in Eq.~\eqref{M_range_result_general} and the necessity of a bare term $\lambda_0\sim\mathcal{O}(10^{-2})$ in the Lagrangian, are general and emerge in all the benchmarks considered.

Note that the 3-loop RGE accuracy is crucial for establishing the condition $\lambda_0\sim\mathcal{O}(10^{-2})$ as, for instance, an analysis restricted to  $n_{RGE}=1$ does admit solutions of Eqns.~\eqref{NO_conditions} with $\lambda_0\simeq 0$~\cite{Brivio:2017dfq}.
This is shown explicitly in Fig.~\ref{plot.num_matching_lam_NH_175}, that reports the matching results for the $\lambda$ parameter obtained with ${\hat m_t = 175}$~GeV. In this case, the $n_{RGE}=1$ dashed curve matches directly the threshold corrections band in the $M_1$ region where Eqn.~\eqref{NO_conditions_m2} is satisfied. This is consistent with what was observed in Ref.~\cite{Brivio:2017dfq}.

\begin{figure}[t]
\hspace*{-1.cm}
\includegraphics[width=1.25\textwidth]{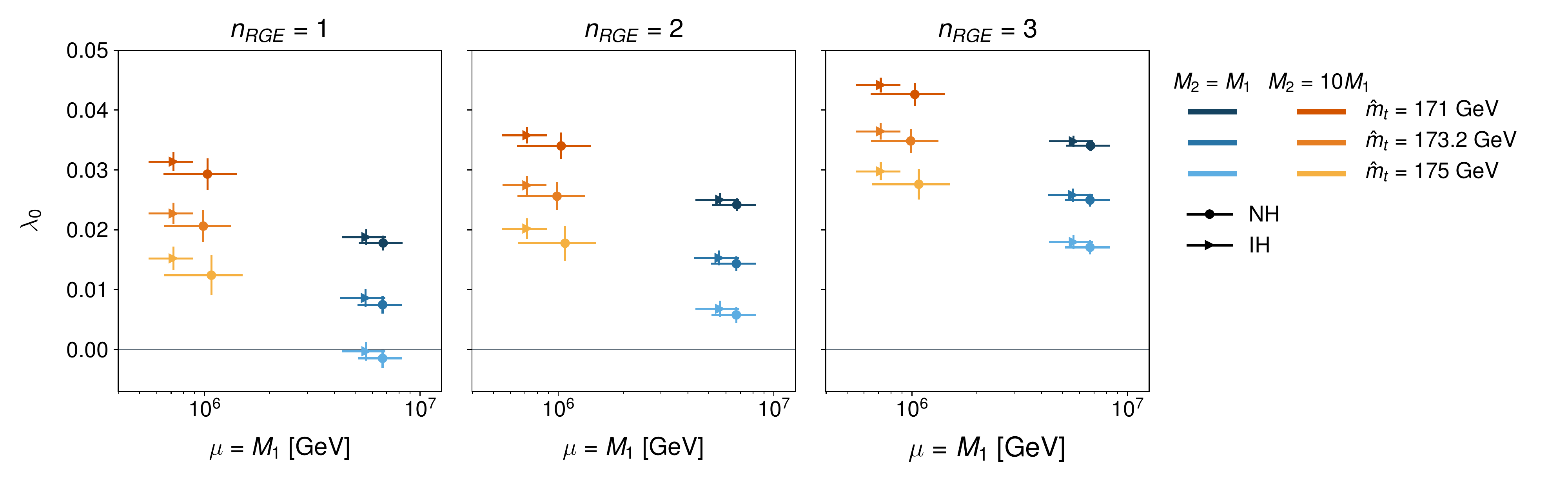}
\caption{Regions of the $(M_1,\lambda_0)$ parameter space in which the neutrino option can be realized, for different choices of: the SM RGE order (panels from left to right), the top mass $\hat m_t$ (colors from lighter to darker), the neutrino mass ordering (dots vs triangles) and $x=M_2/M_1$ (blue for $x=1$, orange for $x=10$).
}\label{plot.logm1_lam}
\end{figure}

The values of ($M_1$, $\lambda_0$) where the neutrino option can be realized for each of the setups considered are summarized in Fig.~\ref{plot.logm1_lam}.
Varying either $n_{RGE}$ or $\hat m_t$ mainly impacts $\lambda(\mu)$, with larger $n_{RGE}$ and smaller $\hat m_t$ giving a smoother running curve and consequently requiring larger values of $\lambda_0$ in the matching.
Choosing the inverse neutrino mass hierarchy rigidly results in slightly lower $M_1$ and slightly larger $\lambda_0$. This is easily understood as follows:
in the IH case the neutrino masses are larger, which leads to larger $\omega_{p\beta}$ (Eqn.~\eqref{CI_def_IH}). The $m^2$ matching relation in Eqn.~\eqref{NO_conditions_m2} selects then a lower $M_1$ region compared to the NH case. Because $\lambda(\mu)$ is larger there, this has the indirect consequence of requiring a larger $\lambda_0$.
Finally, the threshold correction $\Delta m^2(M_1)$ is very sensitive to  the relative size of the Majorana masses $x=M_2/M_1$ (see Fig.~\ref{plot_thresholds_nondegenerate_variation}). Larger $x$ lead to lower $M_1$ being selected and, by the argument above, this indirectly requires larger $\lambda_0$. This explains the relative position of the blue and orange crosses in Fig.~\ref{plot.logm1_lam}.
The correction $\Delta\lambda(M_1)$ does not play any significant role in the numerical analysis as it is always $\ll\lambda(\mu)$ for all the setups considered.

\section{UV embeddings} \label{sec:UV}
The numerical results of Section \ref{sec:numerics} (see also  Ref.~\cite{Brivio:2017dfq,Brdar:2018vjq}) show that a threshold matching defining a boundary condition for $H^\dagger H$
in the seesaw model can be consistent with the lower scale Higgs mass measured experimentally. This conclusion is robust against using $n_{RGE} = \{1,2,3\}$ in the results presented. As the coefficient of $H^\dagger H$ is dimensionful, this modifies the usual concerns of the Electroweak scale Hierarchy problem into an alternate framework.

For this framework to be embedded into a theoretically successful UV completion requires a UV scenario
that can generate the Majorana scale used. This needs to occur in a manner that does not lead to other, larger, threshold matching conditions.
Further, the threshold matching to the $\lambda$ parameter, parametrically $\Delta \lambda \propto \omega^4/16 \pi^2$ can be vanishingly small due to integrating out
the Majorana Neutrino, as $\omega \ll 1$ in order to separate the Majorana mass scale from the effective observed Electroweak scale.
As a direct result, a small matching effect for $\Delta \lambda$ can be subdominant to other UV boundary effects \cite{Brivio:2017dfq},
or even a bare $\lambda_0$ parameter, which is not forbidden by a symmetry.
Explaining the origin of
the required $\lambda(M)$ in a UV scenario would also advance the embedding of this theoretical framework in a more complete UV scenario.

\subsection{The conformal UV embedding of the neutrino option of Ref.~\texorpdfstring{\cite{Brdar:2018vjq}}{[1]}}
Recently a  UV framework for the Majorana scale generation was put forth in Ref.~\cite{Brdar:2018vjq} that addresses most of these theoretical challenges.
The idea is to extend the SM with a set of scalar fields to generate the Majorana scale spontaneously by
satisfying a Gildener-Weinberg \cite{Gildener:1976ih} condition.  The conformal UV completion of the neutrino option
(hereafter the $\mathcal{L}_{CNO}$)  of Ref.~\cite{Brdar:2018vjq} is defined as
\bea\label{CNO}
\mathcal{L}_{CNO} &=& \frac{1}{2} \partial^\mu S \partial_\mu S + \frac{1}{2} \partial^\mu R \partial_\mu R + \overline{N}^p_R i \slashed{\partial} N^p_R - V(H,S,R)  +
\mathcal{L}_{int}, \nn
\mathcal{L}_{int} &=& -\left(\frac{y^{pr}_M}{2} S \,  \overline{N}^p_R \,  N^{r,c}_R
+ \overline{\ell_L^\beta}\tilde H \omega_\beta^{p,\dag}N_p  + h.c. \right), \nn
V(H,S,R) &=& \lambda_S S^4 + \lambda_R R^4 + \lambda_{HS} S^2 (H^\dagger H) + \lambda_{HR} R^2 (H^\dagger H) + \lambda_{SR} S^2 R^2.
\eea
Here $S$,$R$ are real $\rm SU(3)\times SU(2)_L \times U(1)_Y$ singlet scalar fields, and $R$ has an odd charge under a $\mathbb{Z}_2$ symmetry \cite{Brdar:2018vjq}.\footnote{
It is interesting to note the consistency of the field content of this scenario with the the new minimal standard model of Ref.~\cite{Davoudiasl:2004be}.
The latter does not examine a classically scale invariant starting point of parameter space, and is motivated out of minimality
in addressing outstanding experimental deficiencies of the SM. The new minimal standard model does not utilize the neutrino option to generate the Higgs potential
in the parameter space discussed in Ref.~\cite{Davoudiasl:2004be}, but can be considered to be a parameter space variant of the scenario considered here and proposed in
Ref.~\cite{Brdar:2018vjq}.}
The running of the parameters leads to the condition
\bea
\frac{\partial^4 V(H,S,R)}{\partial^4 S} = \lambda_S(\Lambda_{GW}) = 0,
\eea
being satisfied, which leads to the spontaneous breaking of scale invariance once perturbative corrections are included in the
Coleman-Weinberg (CW) potential \cite{Coleman:1973jx}.
$S$ is identified as the pseudo-Goldstone boson of broken scale invariance, the dilaton \cite{Salam:1969rq,Salam:1970qk}.
$S$ also experiences a large breaking of its Goldstone
nature by the coupling $y^{pr}_M$ which has $\mathcal{O}(1)$ entries to the {\it only other fields that are pure singlets}, i.e. $N_R^p$.
The spontaneous breaking of scale invariance gives
\bea\label{massspec}
\langle H^\dagger H \rangle = \langle R \rangle = 0, \quad \quad \langle S \rangle \equiv v_S \neq 0,
\eea
and the mass spectrum is \cite{Gildener:1976ih,Brdar:2018vjq}
\bea
M_N^{pr} = y_M^{pr} \, v_S, \quad \quad M_R = \sqrt{2 \lambda_{SR}} v_S,  \quad \quad M_S = 2 \sqrt{2 \, B} v_S.
\eea
The interaction terms of the seesaw model can be written as in Eqn.~\eqref{L2formulation} with all CP violating phases shifted
to effective couplings and real diagonal entries in a mass matrix. This is conveniently done while by introducing
$N^p = U N^{p'}$ states, defining a rotated diagonal coupling matrix $\tilde{y}_M^{ss} = U^\dagger y_M^{pr} U$.
(The mass matrix in Eqn.~\eqref{L2formulation} is in the diagonal mass basis with the prime superscripts dropped.)
Using this notation the result in Eqn.~\eqref{massspec} explicitly depends on the potential parameters through
\bea
B =  \frac{2 \lambda_{HS}^2 + 2 \lambda_{SR}^2 -  \sum_s (\tilde{y}_M^{ss})^4}{32 \pi^2},
\eea
and it is required that $B>0$ for physical solutions.The remaining contribution to the effective potential is given by  \cite{Gildener:1976ih,Brdar:2018vjq}
\bea
A = \frac{1}{32 \pi^2} \left[2 \lambda_{HS}^2 \left(\log\left[ \lambda_{HS}\right] - \frac{3}{2}\right) + 2 \lambda_{SR}^2 \left(\log\left[ 2 \lambda_{SR}\right] - \frac{3}{2}\right) - \sum_s y_M^4 \left(\log\left[ (y_M^s)^2\right] - \frac{3}{2}\right) \right],
\eea
and the relation between $v_S$ and $\Lambda_{GW}$ is
\bea
\log \left[\frac{v_S}{\Lambda_{GW}}\right] = -\frac{1}{4} - \frac{A}{2B}.
\eea
An additional threshold contribution to $H^\dagger H$ of the form
\bea
\Delta m^2 = \frac{1}{16 \pi^2} \left[  |\omega_p|^2 M_p^2  - \lambda_{HR} M_R^2 \left(1 + 2 \log \frac{M_R^2}{M_S^2}\right)\right],
\eea
is present, and the first term in this expression must dominate for the correct sign to be obtained for $\Delta m^2$. The following set of consistency conditions are required to hold \cite{Brdar:2018vjq}
\bea\label{consistency1}
\frac{2 \lambda_{HS}^2 + 2 \lambda_{SR}^2 -  \sum_s (\tilde{y}_M^{ss})^4}{32 \pi^2} &>& 0,  \nn
  |\omega_p|^2 M_p^2 &>& \lambda_{HR} M_R^2, \nn
|\lambda_{HS}| &<& \frac{1}{16 \pi^2} |\omega_p|^2 M_p^2.
\eea
The parameter space examined in Ref.~\cite{Brdar:2018vjq} is consistent with these conditions and such that
\bea\label{consistency2}
 |\omega_p| &\sim& 10^{-7} - 10^{-3},  \quad \quad y_M \sim \mathcal{O}(0.1),  \quad \quad \Lambda_{GW}[{\rm GeV}] \sim 10^{6} -10^{9}, \nn
 \lambda_{HS} &\sim& 10^{-16}-10^{-19} \quad \quad  \lambda_{SR} \sim \mathcal{O}(0.1) \quad \quad  \lambda_{R} \sim \mathcal{O}(0.1).
\eea
\subsection{Extending the conformal neutrino option}
Define a conformal transformation to be a smooth transformation of the metric
\bea
\tilde{g}_{\mu \, \nu} = \Omega^2(x) g_{\mu \nu},
\eea
that preserves the causal structure of the theory. Further define the conformal weight of the scalar fields of the theory, collectively denoted $\phi$, to be $\tilde{\phi} = \phi/ \Omega(x)$.
A classically conformal UV embedding of the neutrino option requires some further extension due to the existence of gravity. First, this is because the
scalar fields Klein-Gordon equations do not satisfy conformal invariance, until gravitational interactions are included. Second,
the existence of the Planck scale itself is an explicit scale in the complete Lagrangian, calling into question the conformal starting point, and possibly
leading to fine tuning.

Addressing the first challenge is straightforward. The leading interaction terms with
gravity can be considered and fixed to specific classical values. Using a mostly positive metric convention $\{-,+,+,+\}$  the action is given by
\bea\label{fullaction}
S_{CNO} = \int d^4 x \sqrt{-g} \left[-\frac{m_{pl}^2}{2} \, \mathcal{R} - \frac{1}{6} H^\dagger H \, \mathcal{R} \, - \frac{1}{12} S^2 \mathcal{R} - \frac{1}{12} R^2 \mathcal{R} + \mathcal{L}_{SM} + \mathcal{L}_{CNO} + \cdots  \right]. \nn
\eea
Here $\mathcal{R}$ is the Ricci scalar, $m_{pl} = 2.44 \times 10^{18} \, \rm GeV$ is the reduced Planck mass, and $g = \det(g_{\mu \nu})$. Note the scalar kinetic terms flip sign due to the adopted metric convention.
At this stage the theory is necessarily non-renormalizable as the interaction terms $\phi^2 \mathcal{R}$ represent an infinite tower of higher dimensional operators characterizing the
interactions of the graviton with the dynamical field content. Eqn.~\eqref{fullaction} can be studied directly by expanding the metric around flat space in
terms of the dynamical graviton field $h_{\mu \nu}$ as  $g_{\mu \nu} = \eta_{\mu \nu} + h_{\mu \nu}/m_{pl} + \cdots$ which makes it clear that the Lagrangian term
represents a the tower of higher dimensional operators. This set of interactions are dependent on the background field values $\langle \phi^2 \rangle$
and at large field values a large mixing of the scalar degrees of freedom with scalar modes of (non-canonically normalized) gravity results. The identification of the scale $\mu$ in the CW
potential defined in field space, with particle masses leading to the threshold corrections of the neutrino option, ties together running in
field space and the running in energy of the theory, that are formally distinct.

The effects of the non-minimal interactions
with gravity are more easily studied by performing a transformation from the Jordan frame in Eqn.~\eqref{fullaction} to the Einstein frame of the theory. Taking $\tilde{g}_{\mu \nu}  = f(\phi^2) g_{\mu \nu}$ and
\bea\label{secondconfromal}
f(\phi^2) = \left(1 + \frac{H^\dagger H}{3\, m_{pl}^2} + \frac{R^2}{6\, m_{pl}^2} + \frac{S^2}{6\, m_{pl}^2}\right)^{-1}
\eea
as a further conformal transformation takes the theory to canonical form.
The relevant results to study the higher dimensional operators generated already exist in the literature in Refs.~\cite{Burgess:2010zq,Bezrukov:2010jz}.
One finds the expression
\bea\label{jordanframe}
\frac{\mathcal{L}_{CNO}}{\sqrt{- g}} &=& - \frac{m_{pl}^2}{2} \, R - g^{\mu \nu} f^2  \left( (D_\nu H)^\dagger (D_\mu H) + \frac{1}{2} \partial_\nu S \partial_\mu S + \frac{1}{2} \partial_\nu R \partial_\mu R \right)
+ f^2  V(H,S,R), \nn
&\, & + f^2  \overline{N}^p_R i \slashed{\partial} N^p_R + f^2 \, \mathcal{L}_{int} + \cdots.
\eea
We are interested in the possible relation between the values of the parameters in the scalar potential in the CNO and the effects of the higher dimensional operators in
Eqn.~\eqref{jordanframe}. Although expanding in small field values leads to a series of power corrections $\langle S^2\rangle/m_{pl}^2$ to potential terms, setting a lower bound on allowed
$\lambda$ coupling parameter space (when parameter tuning is avoided), we find the values of parameters examined in Ref.~\cite{Brdar:2018vjq} are still viable.

\subsection{Possible fine tuning}

Addressing the consistency of the Planck scale with a conformal embedding of the neutrino option is less straightforward.
On the one hand, the required boundary values of the SM couplings at the Planck scale, including $\lambda$, are expected to be generated in
a consistent UV theory which includes quantum gravity.\footnote{The construction of such a theory is beyond the scope of this work, placing this paper in good company.}
The demand for such an embedding is reinforced by the fact that the SM interactions have Landau poles above the Planck scale. It is natural to speculate
that a UV fixed point can appear in such a theory \cite{Weinberg1978}. If such a conformal field theory in the UV has an interacting fixed point,
then the arguments of Ref.~\cite{Tavares:2013dga}
imply that this scenario can be subject to fine tuning.

Here we review the relevant results of Ref.~\cite{Tavares:2013dga} to make the potential issue clear.
The idea is that a contribution to the scalar two point function will be generated by the transition in the running behavior of the coupling constants of the theory.
Consider the contribution to the scalar two point function determined by an approximately conformal field theory in position space following Ref.~\cite{Tavares:2013dga}
\bea
\langle 0 | T \mathcal{O}^\dagger(x) \mathcal{O}(0) |0 \rangle = \left(\frac{1}{-x^2}\right)^{d-1} \, f(- x^2 \mathcal{N}^2).
\eea
The scale $\mathcal{N}$ is a non perturbative scale characterizing the changing in the running of the coupling constants from the IR free to UV fixed point behavior,
that is assumed. This scale is not associated with a particle mass.
Even so, the key point of Ref.~\cite{Tavares:2013dga} is that by assuming non-analytic dependence on the scale $\mathcal{N}$ in the function $f$, threshold matchings proportional to $\mathcal{N}^2$ are generated
for $H^\dagger H$. We agree that the results of Ref.~\cite{Tavares:2013dga} follow from this assumed non-analytic dependence on $\mathcal{N}$.
As the function $f(- x^2 \mathcal{N}^2)$ is fundamentally non-perturbative it is not possible to draw strong conclusions about its analytic, or non-analytic form, without an
explicit UV theory and non-perturbative study.

In the case at hand, we can examine if sensitivity to the scale $m_{pl}$ for $H^\dagger H$ might already be present using perturbative methods.
The presence of higher dimensional operators in Eqn.~\eqref{jordanframe} indicates this theory is UV incomplete and the presence
of the scale $m_{pl}$ in Eqn~\eqref{fullaction} calls into question the starting assumption of conformal invariance. This theory does have a cut off scale associated with $m_{pl}$.
The cut off scale is background field dependent \cite{Bezrukov:2010jz} but given by $\Lambda \simeq 6 \, m_{pl}$ \cite{Han:2004wt,Burgess:2009ea,Barbon:2009ya}.
The cut off scale comes from unitarity violation generated by the scattering diagram of
the ${\rm SU(2)_L}$ scalar multiplet  coupled to gravity, while the singlet scalar fields do not introduce this cut off dependence \cite{Huggins:1987ea}.
Fundamentally this cut off scale stems from the interaction term $H^\dagger H \mathcal{R}$. The conformal transformation in Eqn.~\eqref{secondconfromal}
used to eliminate  $H^\dagger H \mathcal{R}$ taking the theory to the Einstein frame results in a tower of higher dimensional operators effecting the
CW potential as non-renormalizable classical potential terms.\footnote{The effect of such operators on the effective potential is cogently discussed in Ref.~\cite{Andreassen:2014eha}.}
No $m_{pl}^2$ contribution to $H^\dagger H$ is generated by such operators when expanding the effective potential through the particle thresholds leading to the SM.
An inverse dependence on the scale $1/m_{pl}$ is present breaking conformal symmetry explicitly.

In effective field theory,
non-analytic behavior is usually associated with propagating long distance states of the theory, associated with the poles dictating the properties
of the $S$-matrix. Threshold matchings come about due to fixing that the IR limit of $S$-matrix elements are reproduced when transitioning through a particle threshold.
Non-perturbative matchings can be present, such as the effects of a multi-pole expansion due to underlying structure, which EFT can also be used to represent.
Such effects introduce an inverse dependence on a heavy scale when integrated out, as the multi-pole expansion is a perturbative expansion in ratios of Compton
wavelengths.\footnote{See the SMEFT review \cite{Brivio:2017vri} for more discussion.}

The usual rules of EFT also dictate that a dimension two operator such as $H^\dagger H$
should be considered to have a dimensionful parameter $\propto \Lambda^2 = (6 \, m_{pl})^2$.
If this is the case, then this approach to the Hierarchy problem has severe fine tuning associated with it.
Associating this cut off scale with physical particle masses, one does obtain large contributions to the $H^\dagger H$ operator $\propto \Lambda^2$.
When not associating this cut off scale with physical particle masses, if a threshold contribution to $H^\dagger H$ is still generated, this would be consistent with the arguments in
Ref.~\cite{Tavares:2013dga}.
However, in this case, the cut-off scale is a sign that the effective field theory is smoothly transitioning from a linear SMEFT description to a non-linear EFT description \cite{Burgess:2014lza}
due to the Higgs field mixing with the scalar component of gravity. This mixing introduces non-linearities into the EFT description that require a different description of asymptotic states above and below
the cut off scale.
No particles are integrated out at this scale and
the resulting background field dependent matching contributions across this threshold do not lead to a large shift in the Higgs mass at low field values.
Extrapolating through this scale to large field values is subject to uncertainties due to introduced UV dependence, as the power counting of the EFT breaks down,
but these effects do not necessarily generate a severe fine tuning of the Higgs mass parameter.

\subsection{Comment on exceptional parameter space and Dark Matter}\label{sec:darkmatter}
The set of consistency conditions given in Eqn.~\eqref{consistency1} can be satisfied if alternate parameter space is adopted than considered in
Ref.~\cite{Brdar:2018vjq}. The field content and Lagrangian involving the
$R$ scalar field coupled to the SM is the minimal singlet scalar field Dark Matter model \cite{Silveira:1985rk,McDonald:1993ex,Burgess:2000yq}.
It is interesting to lower the mass of the $R$ scalar to the $\sim \rm GeV$ scale for this reason using such alternate parameter space.  The model parameter space with $\sim {\rm GeV}$ scale exotic scalar states can potentially provide a
successful Dark Matter candidate.
Viable parameter space
for this model's Dark Matter candidate has recently been highly constrained, see Refs.~\cite{Cline:2013gha,Athron:2017kgt,Athron:2018ipf}.
The remaining viable parameter space can be consistent with Eqn.~\eqref{consistency1} at the cost of introducing small, technically natural, scalar
couplings. In particular, the allowed resonance region of parameter space where $2 m_R < m_h$ can be chosen consistent with  the constraints in Refs.~\cite{Cline:2013gha,Athron:2017kgt,Athron:2018ipf}
while a "FIMP" scenario \cite{Hall:2009bx} is present leading to a successful Dark matter relic density satisfying \cite{Bernal:2017kxu}
\bea
\frac{\Omega_R \, h^2}{0.12} = 5.3 \times 10^{21} \ \lambda_{HR}^2 \left(\frac{m_R}{{\rm GeV}} \right).
\eea
In addition, as the scalar $S$ is a pseudo-Goldstone boson, $M_S \ll M_R$;
a scattering channel $R^2 \rightarrow S^2$ is always kinematically open that depletes the relic abundance of $R$. A straightforward calculation yields
\bea
\sigma_{ann} v_{rel} \sim \frac{\lambda_{SR}^2}{4 \pi M_R^2} \sim \frac{1}{16 \pi v_S^2}.
\eea
For the exceptional parameter space to be viable, it is also necessary that additional Gildener-Weinberg \cite{Gildener:1976ih} conditions discussed in Ref.~\cite{Brdar:2018vjq}
must not be satisfied, to avoid spontaneously breaking scale invariance at scales other than $\Lambda_{GW}$.
Finally, due to the presence of the scalar $S$, and the self interactions of the $R$ field, the viability of the model is dependent on a nontrivial thermal history
and also the balance of freeze-in and freeze out effects. We leave a detailed investigation of this possibility to a future publication.

\section{Conclusions} \label{sec:conclusions}
In this paper, we have examined the neutrino option, where the Electroweak scale is generated simultaneously with neutrino masses.
We have examined the numerical consistency of this scenario using one, two and three loop RGE equations for the SM, and one loop running of the Weinberg operator.
We have developed a consistent NLO framework for such studies by determining the full set of ($\propto \omega^2$) one loop corrections to the leading tree level matching.
We have identified the requirement of a $\lambda$ parameter $\sim \mathcal{O}(10^{-2})$ at the matching scale
and confirmed previous results that indicate that this scenario predicts Majorana states around the scales
$5\cdot \unit[10^5]{GeV}
\;\lesssim\; M_1 \;\lesssim\; \unit[10^7]{GeV}$.
We have also extended the conformal neutrino option scenario of Brdar et.al.  \cite{Brdar:2018vjq} to include the leading couplings to gravity. The conformal neutrino option necessarily includes
a cut off scale due to scalar-graviton mixing. Nevertheless, we have also argued that this cut off scale
does not result in a large fine tuning unless it is an indication of a UV-completion with Planck scale states, and in addition this cut off scale can be directly interpretted as a sign of a
transition to a non-linear EFT set up around the Planck scale with no new states. It is far from obvious
that calculable Planck scale threshold corrections to the Higgs mass can be demonstrated in this case.
Overall, our results strongly support further investigations of the neutrino option and its conformal UV embedding being pursued.

Extensions to the Standard Model motivated out of the experimental
requirement of neutrino mass generation remarkably remain an option to address the Hierarchy problem.

\begin{acknowledgments}
MT and IB acknowledge generous support from the Villum Fonden and partial support by the Danish National Research Foundation (DNRF91) through the Discovery centre.
We thank  E. Bjerrum-Bohr, J. Bourjaily, J. Cline, A. Denner, W. Goldberger, H. Haber,  A. Manohar, H. Murayama, S. Patil, P. Perez, S. Petcov, W. Skiba and J. Valle for useful discussions and comments.
\end{acknowledgments}

\appendix
\section{Evaluating correlation functions in alternate Seesaw model formulations}
Evaluating correlation functions when examining the neutrino option can be subtle due to the presence of a Majorana field.
Feynman rules for Majorana particles exist in Ref.~\cite{Denner:1992vza,Denner:1992me,Dreiner:2008tw}, but subtleties can still remain, as we illustrate in this Appendix.
Consider evaluating a vacuum expectation value of a correlation function using Wick's theorem \cite{PhysRev.80.268}
and comparing the results using the Lagrangians $\mathcal{L}_N$, $\mathcal{L}'_N$.
The interaction terms of $\mathcal{L}_N$, $\mathcal{L}'_N$ are classically identical,
\bea
\frac{1}{2}\Bigg[ \overline{\ell_L^\beta}\tilde H \omega_\beta^{p,\dag}N_p + \overline{\ell_L^{c\beta}}\tilde H^* \omega_\beta^{p,T}N_p
 +\overline{N_p} \omega_\beta^{p,*}\tilde H ^T \ell_L^{c\beta} +\overline{N_p} \omega_\beta^p\tilde H^\dag \ell_L^{\beta}
 \Bigg] \equiv \Bigg[ \overline{\ell_L^\beta}\tilde H \omega_\beta^{p,\dag}N_p  +\overline{N_p} \omega_\beta^p\tilde H^\dag \ell_L^{\beta}
 \Bigg]. \nonumber
\eea
Now consider evaluating $\Pi_{HH}$  using $\mathcal{L}'_N$. Wick's theorem gives
\bea
i \Pi_{HH^\dagger} &=&\frac{1}{2!} \langle 0 | T \{ H^\dagger(x) H(y)  \int d z^4 \mathcal{L}'_N(z)  \int d w^4 \mathcal{L}'_N(w)\}| 0 \rangle, \nn
&=&
\acontraction{\langle 0| T H^\dagger}{H}{\overline{\ell_L^\beta} \tilde H N_p  \overline{N_p} }{\tilde H^\dagger}
\acontraction[1.5ex]{\langle 0| T H^\dagger H \overline{\ell_L^\beta} \tilde H}{N_p}{}{\overline{N_p}}
\bcontraction{\langle 0| T }{H^\dagger}{H \overline{\ell_L^\beta}}{\tilde H}
\bcontraction[1.5ex] {\langle 0| T H^\dagger H}{\overline{\ell_L^\beta}}{ \tilde H  N_p  \overline{N_p}  \tilde H^\dagger}{\ell_L^\beta}
\langle 0| T H^\dagger H \overline{\ell_L^\beta} \tilde H  N_p  \overline{N_p}  \tilde H^\dagger \ell_L^\beta | 0 \rangle |\omega_p|^2, \nn
&=& - \frac{i \, |\omega_p|^2M_p^2}{8\pi^2} \left(1+\log\frac{\mu^2}{M_p^2}\right).
\eea
Due to the presence of a Majorana field, one can use the transformation
\bea
 \overline{\ell_L^\beta}\tilde H \omega_\beta^{p,\dag}N_p   \rightarrow  \overline{N_p} \omega_\beta^{p,\star} \tilde H^T \ell_L^{c ,\beta}, \quad \quad
 \overline{N_p} \omega_\beta^p\tilde H^\dagger \ell_L^{\beta} \rightarrow  \overline{\ell_L^{c, \beta}} (\omega_\beta^p)^T \tilde H^\star N_p,
\eea
before the contractions in Wick's theorem are evaluated when considering the  "self square" interaction terms  $[ \overline{\ell_L^\beta}\tilde H \omega_\beta^{p,\dag}N_p]^2$,
$[\overline{N_p} \omega_\beta^p\tilde H^\dag \ell_L^{\beta}]^2$, also present in the time-ordered exponential. These transformations
follow from transposing the interaction Lagrangian, and the Fermi-statistics of the fermionic field operators. Transposition of the Lagrangian is justified
as the individual Lagrangian terms are invariant under the (compact Euclideanized $\rm SO(4) \simeq SU(2)_L \times SU(2)_R$) Lorentz group, and the global
$\rm SU(3) \times SU(2)_L \times U(1)_Y$.
Although this transformation still allows the Majorana field $N$ to be contracted defining a two point function, as the correlation functions
\begin{align*}
\langle 0 | \ell^c \, \overline{\ell} | 0 \rangle &= 0, \quad \langle 0 | \overline{\ell^c} \, \ell | 0 \rangle = 0,  \quad
\langle 0 | \ell \, \overline{\ell}^c | 0 \rangle = 0, \quad \langle 0 | \overline{\ell} \, \ell^c | 0 \rangle = 0,
\end{align*}
no additional contribution to the sum of the time ordered product in Wick's theorem results. These allowed transformations do lead to
further contractions when considering other matrix elements, such as the three point and four point functions evaluated in this work, when using $\mathcal{L}'_N$.
A careful use of the Feynman rules of Refs.~\cite{Denner:1992vza,Denner:1992me} still evaluates the correlation function in each case
only using non-zero correlation functions $\acontraction[0.5ex]{}{\ell}{}{\bar{\ell}}  \ell \bar{\ell}$, $\acontraction[0.5ex]{}{\ell^c}{}{\bar{\ell}^c}  \ell^c \bar{\ell^c}$.

On the other hand, using $\mathcal{L}_N$ and neglecting such "self square" interaction terms, directly gives
\bea
i \Pi_{HH^\dagger} &=& - \frac{i \, |\omega_p|^2M_p^2}{16\pi^2} \left(1+\log\frac{\mu^2}{M_p^2}\right).
\eea
In this case, the "self square" of the manifestly charge symmetric  interaction Lagrangian terms  include
\bea
\Bigg[ \overline{\ell_L^\beta}\tilde H \omega_\beta^{p,\dag}N_p + \overline{\ell_L^{c\beta}}\tilde H^* \omega_\beta^{p,T}N_p \Bigg]^2.
\eea
Using the charge conjugation identities, transposing the total Lagrangian, and Fermi-statistics gives
\bea
\Bigg[ \overline{\ell_L^\beta}\tilde H \omega_\beta^{p,\dag}N_p + \overline{\ell_L^{c\beta}}\tilde H^* \omega_\beta^{p,T}N_p \Bigg]^2 =
\Bigg[ \overline{\ell_L^\beta}\tilde H \omega_\beta^{p,\dag}N_p + \overline{\ell_L^{c\beta}}\tilde H^* \omega_\beta^{p,T}N_p \Bigg]
\Bigg[\overline{N_p} \omega_\beta^{p,*}\tilde H ^T \ell_L^{c\beta} +\overline{N_p} \omega_\beta^p\tilde H^\dag \ell_L^{\beta}\Bigg]. \nonumber
\eea
Using a manifestly charge symmetric interaction Lagrangian results in the additional contributions of this form, due to these allowed manipulations when evaluating
the Wick expansion. An additional overall factor of two results evaluating $i \Pi_{HH}$, which brings the results obtained
using $\mathcal{L}_N$, $\mathcal{L}'_N$ into agreement. A sign of the need to perform transformations of this form in evaluating
the Wick expansion is the presence of non-zero contributions
\bea
\langle 0 | \ell^c(x) \,\ell(y) | 0 \rangle, \quad \langle 0 | \ell(y) \,\ell^c(x) | 0 \rangle \neq 0.
\eea
These terms do not have a simple expression in terms of a charge conjugation matrix and gamma matrices, but indicate a transposition is required to evaluate the spin sum
in some terms in the Wick expansion. On the other hand, when considering the three point function, the presence of the $\ell$ and $\ell^c$ fields leads to a direct evaluation of
the Wick expansion in the  $\mathcal{L}_N$ formulation. Judiciously utilizing the transposition transformations can simplify various calculations when using $\mathcal{L}_N$ or $\mathcal{L}_N'$.

\bibliographystyle{JHEP}
\bibliography{bibliography2}

\providecommand{\href}[2]{#2}\begingroup\raggedright\begin{thebibliography}{10}

\bibitem{Brdar:2018vjq}
V.~Brdar, Y.~Emonds, A.~J. Helmboldt, and M.~Lindner, {\it {The Conformal UV
  Completion of the Neutrino Option}},
  \href{http://arxiv.org/abs/1807.11490}{{\tt arXiv:1807.11490}}.

\bibitem{Minkowski:1977sc}
P.~Minkowski, {\it {$\mu \to e\gamma$ at a Rate of One Out of $10^{9}$ Muon
  Decays?}},  {\em Phys. Lett.} {\bf B67} (1977) 421--428.

\bibitem{GellMann:1980vs}
M.~Gell-Mann, P.~Ramond, and R.~Slansky, {\it {Complex Spinors and Unified
  Theories}},  {\em Conf. Proc.} {\bf C790927} (1979) 315--321,
  [\href{http://arxiv.org/abs/1306.4669}{{\tt arXiv:1306.4669}}].

\bibitem{Mohapatra:1979ia}
R.~N. Mohapatra and G.~Senjanovic, {\it {Neutrino Mass and Spontaneous Parity
  Violation}},  {\em Phys. Rev. Lett.} {\bf 44} (1980) 912.

\bibitem{Yanagida:1980xy}
T.~Yanagida, {\it {Horizontal Symmetry and Masses of Neutrinos}},  {\em Prog.
  Theor. Phys.} {\bf 64} (1980) 1103.

\bibitem{Schechter:1980gr}
J.~Schechter and J.~W.~F. Valle, {\it {Neutrino Masses in SU(2) x U(1)
  Theories}},  {\em Phys. Rev.} {\bf D22} (1980) 2227.

\bibitem{Vissani:1997ys}
F.~Vissani, {\it {Do experiments suggest a hierarchy problem?}},  {\em Phys.
  Rev.} {\bf D57} (1998) 7027--7030,
  [\href{http://arxiv.org/abs/hep-ph/9709409}{{\tt hep-ph/9709409}}].

\bibitem{Brivio:2017dfq}
I.~Brivio and M.~Trott, {\it {Radiatively Generating the Higgs Potential and
  Electroweak Scale via the Seesaw Mechanism}},  {\em Phys. Rev. Lett.} {\bf
  119} (2017), no.~14 141801, [\href{http://arxiv.org/abs/1703.10924}{{\tt
  arXiv:1703.10924}}].

\bibitem{deGouvea:2014lva}
A.~de~Gouvea, J.~Herrero-Garcia, and A.~Kobach, {\it {Neutrino Masses, Grand
  Unification, and Baryon Number Violation}},  {\em Phys. Rev.} {\bf D90}
  (2014), no.~1 016011, [\href{http://arxiv.org/abs/1404.4057}{{\tt
  arXiv:1404.4057}}].

\bibitem{Kobach:2016ami}
A.~Kobach, {\it {Baryon Number, Lepton Number, and Operator Dimension in the
  Standard Model}},  {\em Phys. Lett.} {\bf B758} (2016) 455--457,
  [\href{http://arxiv.org/abs/1604.05726}{{\tt arXiv:1604.05726}}].

\bibitem{Mohapatra:1986aw}
R.~N. Mohapatra, {\it {Mechanism for Understanding Small Neutrino Mass in
  Superstring Theories}},  {\em Phys. Rev. Lett.} {\bf 56} (1986) 561--563.

\bibitem{Mohapatra:1986bd}
R.~N. Mohapatra and J.~W.~F. Valle, {\it {Neutrino Mass and Baryon Number
  Nonconservation in Superstring Models}},  {\em Phys. Rev.} {\bf D34} (1986)
  1642. [,235(1986)].

\bibitem{Bernabeu:1987gr}
J.~Bernabeu, A.~Santamaria, J.~Vidal, A.~Mendez, and J.~W.~F. Valle, {\it
  {Lepton Flavor Nonconservation at High-Energies in a Superstring Inspired
  Standard Model}},  {\em Phys. Lett.} {\bf B187} (1987) 303--308.

\bibitem{Davoudiasl:2014pya}
H.~Davoudiasl and I.~M. Lewis, {\it {Right-Handed Neutrinos as the Origin of
  the Electroweak Scale}},  {\em Phys. Rev.} {\bf D90} (2014), no.~3 033003,
  [\href{http://arxiv.org/abs/1404.6260}{{\tt arXiv:1404.6260}}].

\bibitem{Casas:1998cf}
J.~A. Casas, V.~Di~Clemente, and M.~Quiros, {\it {The Effective potential in
  the presence of several mass scales}},  {\em Nucl. Phys.} {\bf B553} (1999)
  511--530, [\href{http://arxiv.org/abs/hep-ph/9809275}{{\tt hep-ph/9809275}}].

\bibitem{Casas:1999cd}
J.~A. Casas, V.~Di~Clemente, A.~Ibarra, and M.~Quiros, {\it {Massive neutrinos
  and the Higgs mass window}},  {\em Phys. Rev.} {\bf D62} (2000) 053005,
  [\href{http://arxiv.org/abs/hep-ph/9904295}{{\tt hep-ph/9904295}}].

\bibitem{Bambhaniya:2016rbb}
G.~Bambhaniya, P.~Bhupal~Dev, S.~Goswami, S.~Khan, and W.~Rodejohann, {\it
  {Naturalness, Vacuum Stability and Leptogenesis in the Minimal Seesaw
  Model}},  {\em Phys. Rev.} {\bf D95} (2017), no.~9 095016,
  [\href{http://arxiv.org/abs/1611.03827}{{\tt arXiv:1611.03827}}].

\bibitem{Callan:1970ze}
C.~G. Callan, Jr., S.~R. Coleman, and R.~Jackiw, {\it {A New improved energy -
  momentum tensor}},  {\em Annals Phys.} {\bf 59} (1970) 42--73.

\bibitem{Coleman:1970je}
S.~R. Coleman and R.~Jackiw, {\it {Why dilatation generators do not generate
  dilatations?}},  {\em Annals Phys.} {\bf 67} (1971) 552--598.

\bibitem{Bardeen:1995kv}
W.~A. Bardeen, {\it {On naturalness in the standard model}},  in {\em {Ontake
  Summer Institute on Particle Physics Ontake Mountain, Japan, August
  27-September 2, 1995}}, 1995.

\bibitem{Capper:1974ic}
D.~M. Capper and M.~J. Duff, {\it {Trace anomalies in dimensional
  regularization}},  {\em Nuovo Cim.} {\bf A23} (1974) 173--183.

\bibitem{Bellazzini:2014yua}
B.~Bellazzini, C.~Csaki, and J.~Serra, {\it {Composite Higgses}},  {\em Eur.
  Phys. J.} {\bf C74} (2014), no.~5 2766,
  [\href{http://arxiv.org/abs/1401.2457}{{\tt arXiv:1401.2457}}].

\bibitem{Buchmuller:1985jz}
W.~Buchm\"uller and D.~Wyler, {\it {Effective Lagrangian Analysis of New
  Interactions and Flavor Conservation}},  {\em Nucl.Phys.} {\bf B268} (1986)
  621--653.

\bibitem{Grzadkowski:2010es}
B.~Grzadkowski, M.~Iskrzynski, M.~Misiak, and J.~Rosiek, {\it {Dimension-Six
  Terms in the Standard Model Lagrangian}},  {\em JHEP} {\bf 1010} (2010) 085,
  [\href{http://arxiv.org/abs/1008.4884}{{\tt arXiv:1008.4884}}].

\bibitem{Brivio:2017vri}
I.~Brivio and M.~Trott, {\it {The Standard Model as an Effective Field
  Theory}},  \href{http://arxiv.org/abs/1706.08945}{{\tt arXiv:1706.08945}}.

\bibitem{Weinberg:1979sa}
S.~Weinberg, {\it {Baryon and Lepton Nonconserving Processes}},  {\em
  Phys.Rev.Lett.} {\bf 43} (1979) 1566--1570.

\bibitem{Wilczek:1979hc}
F.~Wilczek and A.~Zee, {\it {Operator Analysis of Nucleon Decay}},  {\em Phys.
  Rev. Lett.} {\bf 43} (1979) 1571--1573.

\bibitem{Broncano:2002rw}
A.~Broncano, M.~B. Gavela, and E.~E. Jenkins, {\it {The Effective Lagrangian
  for the seesaw model of neutrino mass and leptogenesis}},  {\em Phys. Lett.}
  {\bf B552} (2003) 177--184, [\href{http://arxiv.org/abs/hep-ph/0210271}{{\tt
  hep-ph/0210271}}]. [Erratum: Phys. Lett.B636,332(2006)].

\bibitem{Elgaard-Clausen:2017xkq}
G.~Elgaard-Clausen and M.~Trott, {\it {On expansions in neutrino effective
  field theory}},  {\em JHEP} {\bf 11} (2017) 088,
  [\href{http://arxiv.org/abs/1703.04415}{{\tt arXiv:1703.04415}}].

\bibitem{Majorana:1937vz}
E.~Majorana, {\it {Teoria simmetrica dell'elettrone e del positrone}},  {\em
  Nuovo Cim.} {\bf 14} (1937) 171--184.

\bibitem{Bilenky:1980cx}
S.~M. Bilenky, J.~Hosek, and S.~T. Petcov, {\it {On Oscillations of Neutrinos
  with Dirac and Majorana Masses}},  {\em Phys. Lett.} {\bf 94B} (1980)
  495--498.

\bibitem{Pontecorvo:1957cp}
B.~Pontecorvo, {\it {Mesonium and anti-mesonium}},  {\em Sov. Phys. JETP} {\bf
  6} (1957) 429. [Zh. Eksp. Teor. Fiz.33,549(1957)].

\bibitem{Maki:1962mu}
Z.~Maki, M.~Nakagawa, and S.~Sakata, {\it {Remarks on the unified model of
  elementary particles}},  {\em Prog. Theor. Phys.} {\bf 28} (1962) 870--880.

\bibitem{Gonzalez-Garcia:2014bfa}
M.~C. Gonzalez-Garcia, M.~Maltoni, and T.~Schwetz, {\it {Updated fit to three
  neutrino mixing: status of leptonic CP violation}},  {\em JHEP} {\bf 11}
  (2014) 052, [\href{http://arxiv.org/abs/1409.5439}{{\tt arXiv:1409.5439}}].

\bibitem{Esteban:2016qun}
I.~Esteban, M.~C. Gonzalez-Garcia, M.~Maltoni, I.~Martinez-Soler, and
  T.~Schwetz, {\it {Updated fit to three neutrino mixing: exploring the
  accelerator-reactor complementarity}},  {\em JHEP} {\bf 01} (2017) 087,
  [\href{http://arxiv.org/abs/1611.01514}{{\tt arXiv:1611.01514}}].

\bibitem{Casas:2001sr}
J.~A. Casas and A.~Ibarra, {\it {Oscillating neutrinos and muon ---> e,
  gamma}},  {\em Nucl. Phys.} {\bf B618} (2001) 171--204,
  [\href{http://arxiv.org/abs/hep-ph/0103065}{{\tt hep-ph/0103065}}].

\bibitem{Ibarra:2011xn}
A.~Ibarra, E.~Molinaro, and S.~T. Petcov, {\it {Low Energy Signatures of the
  TeV Scale See-Saw Mechanism}},  {\em Phys. Rev.} {\bf D84} (2011) 013005,
  [\href{http://arxiv.org/abs/1103.6217}{{\tt arXiv:1103.6217}}].

\bibitem{Grimus:1989pu}
W.~Grimus and H.~Neufeld, {\it {Radiative Neutrino Masses in an SU(2) X U(1)
  Model}},  {\em Nucl. Phys.} {\bf B325} (1989) 18--32.

\bibitem{Grimus:2002nk}
W.~Grimus and L.~Lavoura, {\it {One-loop corrections to the seesaw mechanism in
  the multi-Higgs-doublet standard model}},  {\em Phys. Lett.} {\bf B546}
  (2002) 86--95, [\href{http://arxiv.org/abs/hep-ph/0207229}{{\tt
  hep-ph/0207229}}].

\bibitem{Dev:2012sg}
P.~S.~B. Dev and A.~Pilaftsis, {\it {Minimal Radiative Neutrino Mass Mechanism
  for Inverse Seesaw Models}},  {\em Phys. Rev.} {\bf D86} (2012) 113001,
  [\href{http://arxiv.org/abs/1209.4051}{{\tt arXiv:1209.4051}}].

\bibitem{Fernandez-Martinez:2015hxa}
E.~Fernandez-Martinez, J.~Hernandez-Garcia, J.~Lopez-Pavon, and M.~Lucente,
  {\it {Loop level constraints on Seesaw neutrino mixing}},  {\em JHEP} {\bf
  10} (2015) 130, [\href{http://arxiv.org/abs/1508.03051}{{\tt
  arXiv:1508.03051}}].

\bibitem{Buttazzo:2013uya}
D.~Buttazzo, G.~Degrassi, P.~P. Giardino, G.~F. Giudice, F.~Sala, A.~Salvio,
  and A.~Strumia, {\it {Investigating the near-criticality of the Higgs
  boson}},  {\em JHEP} {\bf 12} (2013) 089,
  [\href{http://arxiv.org/abs/1307.3536}{{\tt arXiv:1307.3536}}].

\bibitem{Babu:1993qv}
K.~S. Babu, C.~N. Leung, and J.~T. Pantaleone, {\it {Renormalization of the
  neutrino mass operator}},  {\em Phys. Lett.} {\bf B319} (1993) 191--198,
  [\href{http://arxiv.org/abs/hep-ph/9309223}{{\tt hep-ph/9309223}}].

\bibitem{Antusch:2001ck}
S.~Antusch, M.~Drees, J.~Kersten, M.~Lindner, and M.~Ratz, {\it {Neutrino mass
  operator renormalization revisited}},  {\em Phys. Lett.} {\bf B519} (2001)
  238--242, [\href{http://arxiv.org/abs/hep-ph/0108005}{{\tt hep-ph/0108005}}].

\bibitem{Casas:1999tg}
J.~A. Casas, J.~R. Espinosa, A.~Ibarra, and I.~Navarro, {\it {General RG
  equations for physical neutrino parameters and their phenomenological
  implications}},  {\em Nucl. Phys.} {\bf B573} (2000) 652--684,
  [\href{http://arxiv.org/abs/hep-ph/9910420}{{\tt hep-ph/9910420}}].

\bibitem{Pilaftsis:2003gt}
A.~Pilaftsis and T.~E.~J. Underwood, {\it {Resonant leptogenesis}},  {\em Nucl.
  Phys.} {\bf B692} (2004) 303--345,
  [\href{http://arxiv.org/abs/hep-ph/0309342}{{\tt hep-ph/0309342}}].

\bibitem{Casas:2010wm}
J.~A. Casas, J.~M. Moreno, N.~Rius, R.~Ruiz~de Austri, and B.~Zaldivar, {\it
  {Fair scans of the seesaw. Consequences for predictions on LFV processes}},
  {\em JHEP} {\bf 03} (2011) 034, [\href{http://arxiv.org/abs/1010.5751}{{\tt
  arXiv:1010.5751}}].

\bibitem{Gildener:1976ih}
E.~Gildener and S.~Weinberg, {\it {Symmetry Breaking and Scalar Bosons}},  {\em
  Phys. Rev.} {\bf D13} (1976) 3333.

\bibitem{Davoudiasl:2004be}
H.~Davoudiasl, R.~Kitano, T.~Li, and H.~Murayama, {\it {The New minimal
  standard model}},  {\em Phys. Lett.} {\bf B609} (2005) 117--123,
  [\href{http://arxiv.org/abs/hep-ph/0405097}{{\tt hep-ph/0405097}}].

\bibitem{Coleman:1973jx}
S.~R. Coleman and E.~J. Weinberg, {\it {Radiative Corrections as the Origin of
  Spontaneous Symmetry Breaking}},  {\em Phys. Rev.} {\bf D7} (1973)
  1888--1910.

\bibitem{Salam:1969rq}
A.~Salam and J.~A. Strathdee, {\it {Nonlinear realizations. 1: The Role of
  Goldstone bosons}},  {\em Phys. Rev.} {\bf 184} (1969) 1750--1759.

\bibitem{Salam:1970qk}
A.~Salam and J.~A. Strathdee, {\it {Nonlinear realizations. 2. Conformal
  symmetry}},  {\em Phys. Rev.} {\bf 184} (1969) 1760--1768.

\bibitem{Burgess:2010zq}
C.~P. Burgess, H.~M. Lee, and M.~Trott, {\it {Comment on Higgs Inflation and
  Naturalness}},  {\em JHEP} {\bf 07} (2010) 007,
  [\href{http://arxiv.org/abs/1002.2730}{{\tt arXiv:1002.2730}}].

\bibitem{Bezrukov:2010jz}
F.~Bezrukov, A.~Magnin, M.~Shaposhnikov, and S.~Sibiryakov, {\it {Higgs
  inflation: consistency and generalisations}},  {\em JHEP} {\bf 01} (2011)
  016, [\href{http://arxiv.org/abs/1008.5157}{{\tt arXiv:1008.5157}}].

\bibitem{Weinberg1978}
S.~Weinberg, {\em Critical Phenomena for Field Theorists}, pp.~1--52.
\newblock Springer US, Boston, MA, 1978.

\bibitem{Tavares:2013dga}
G.~Marques~Tavares, M.~Schmaltz, and W.~Skiba, {\it {Higgs mass naturalness and
  scale invariance in the UV}},  {\em Phys. Rev.} {\bf D89} (2014), no.~1
  015009, [\href{http://arxiv.org/abs/1308.0025}{{\tt arXiv:1308.0025}}].

\bibitem{Han:2004wt}
T.~Han and S.~Willenbrock, {\it {Scale of quantum gravity}},  {\em Phys. Lett.}
  {\bf B616} (2005) 215--220, [\href{http://arxiv.org/abs/hep-ph/0404182}{{\tt
  hep-ph/0404182}}].

\bibitem{Burgess:2009ea}
C.~P. Burgess, H.~M. Lee, and M.~Trott, {\it {Power-counting and the Validity
  of the Classical Approximation During Inflation}},  {\em JHEP} {\bf 09}
  (2009) 103, [\href{http://arxiv.org/abs/0902.4465}{{\tt arXiv:0902.4465}}].

\bibitem{Barbon:2009ya}
J.~L.~F. Barbon and J.~R. Espinosa, {\it {On the Naturalness of Higgs
  Inflation}},  {\em Phys. Rev.} {\bf D79} (2009) 081302,
  [\href{http://arxiv.org/abs/0903.0355}{{\tt arXiv:0903.0355}}].

\bibitem{Huggins:1987ea}
S.~R. Huggins and D.~J. Toms, {\it {One Graviton Exchange Interaction of
  Nonminimally Coupled Scalar Fields}},  {\em Class. Quant. Grav.} {\bf 4}
  (1987) 1509.

\bibitem{Andreassen:2014eha}
A.~Andreassen, W.~Frost, and M.~D. Schwartz, {\it {Consistent Use of Effective
  Potentials}},  {\em Phys. Rev.} {\bf D91} (2015), no.~1 016009,
  [\href{http://arxiv.org/abs/1408.0287}{{\tt arXiv:1408.0287}}].

\bibitem{Burgess:2014lza}
C.~P. Burgess, S.~P. Patil, and M.~Trott, {\it {On the Predictiveness of
  Single-Field Inflationary Models}},  {\em JHEP} {\bf 06} (2014) 010,
  [\href{http://arxiv.org/abs/1402.1476}{{\tt arXiv:1402.1476}}].

\bibitem{Silveira:1985rk}
V.~Silveira and A.~Zee, {\it {Scalar phantoms}},  {\em Phys. Lett.} {\bf 161B}
  (1985) 136--140.

\bibitem{McDonald:1993ex}
J.~McDonald, {\it {Gauge singlet scalars as cold dark matter}},  {\em Phys.
  Rev.} {\bf D50} (1994) 3637--3649,
  [\href{http://arxiv.org/abs/hep-ph/0702143}{{\tt hep-ph/0702143}}].

\bibitem{Burgess:2000yq}
C.~P. Burgess, M.~Pospelov, and T.~ter Veldhuis, {\it {The Minimal model of
  nonbaryonic dark matter: A Singlet scalar}},  {\em Nucl. Phys.} {\bf B619}
  (2001) 709--728, [\href{http://arxiv.org/abs/hep-ph/0011335}{{\tt
  hep-ph/0011335}}].

\bibitem{Cline:2013gha}
J.~M. Cline, K.~Kainulainen, P.~Scott, and C.~Weniger, {\it {Update on scalar
  singlet dark matter}},  {\em Phys. Rev.} {\bf D88} (2013) 055025,
  [\href{http://arxiv.org/abs/1306.4710}{{\tt arXiv:1306.4710}}]. [Erratum:
  Phys. Rev.D92,no.3,039906(2015)].

\bibitem{Athron:2017kgt}
{\bf GAMBIT} Collaboration, P.~Athron et~al., {\it {Status of the scalar
  singlet dark matter model}},  {\em Eur. Phys. J.} {\bf C77} (2017), no.~8
  568, [\href{http://arxiv.org/abs/1705.07931}{{\tt arXiv:1705.07931}}].

\bibitem{Athron:2018ipf}
P.~Athron, J.~M. Cornell, F.~Kahlhoefer, J.~Mckay, P.~Scott, and S.~Wild, {\it
  {Impact of vacuum stability, perturbativity and XENON1T on global fits of
  $\mathbb{Z}_2$ and $\mathbb{Z}_3$ scalar singlet dark matter}},
  \href{http://arxiv.org/abs/1806.11281}{{\tt arXiv:1806.11281}}.

\bibitem{Hall:2009bx}
L.~J. Hall, K.~Jedamzik, J.~March-Russell, and S.~M. West, {\it {Freeze-In
  Production of FIMP Dark Matter}},  {\em JHEP} {\bf 03} (2010) 080,
  [\href{http://arxiv.org/abs/0911.1120}{{\tt arXiv:0911.1120}}].

\bibitem{Bernal:2017kxu}
N.~Bernal, M.~Heikinheimo, T.~Tenkanen, K.~Tuominen, and V.~Vaskonen, {\it {The
  Dawn of FIMP Dark Matter: A Review of Models and Constraints}},  {\em Int. J.
  Mod. Phys.} {\bf A32} (2017), no.~27 1730023,
  [\href{http://arxiv.org/abs/1706.07442}{{\tt arXiv:1706.07442}}].

\bibitem{Denner:1992vza}
A.~Denner, H.~Eck, O.~Hahn, and J.~Kublbeck, {\it {Feynman rules for fermion
  number violating interactions}},  {\em Nucl. Phys.} {\bf B387} (1992)
  467--481.

\bibitem{Denner:1992me}
A.~Denner, H.~Eck, O.~Hahn, and J.~Kublbeck, {\it {Compact Feynman rules for
  Majorana fermions}},  {\em Phys. Lett.} {\bf B291} (1992) 278--280.

\bibitem{Dreiner:2008tw}
H.~K. Dreiner, H.~E. Haber, and S.~P. Martin, {\it {Two-component spinor
  techniques and Feynman rules for quantum field theory and supersymmetry}},
  {\em Phys. Rept.} {\bf 494} (2010) 1--196,
  [\href{http://arxiv.org/abs/0812.1594}{{\tt arXiv:0812.1594}}].

\bibitem{PhysRev.80.268}
G.~C. Wick, {\it {The Evaluation of the Collision Matrix}},  {\em Phys. Rev.}
  {\bf 80} (1950) 268--272. [,592(1950)].

\end{thebibliography}\endgroup
\end{document}